\pdfoutput=1

\documentclass[aps,preprint,groupedaddress]{revtex4-1} %use this
\usepackage{amsmath}
\usepackage{graphicx}% Include figure files
\usepackage[lofdepth,lotdepth,caption=false]{subfig}

\usepackage{color}

\usepackage{algorithmic}

\usepackage[colorlinks=true,
            urlcolor=blue,
            citecolor=blue,
            linkcolor=black,
            pdfstartview=FitH,
            pdfpagemode=None]{hyperref}

\usepackage{empheq}
\usepackage{cases}

\def\ket#1{\mathinner{|{#1}\rangle}}

\newcommand{\ketbra}[2]{|~#1~\rangle~\langle#2|}

\newcommand{\up}{\uparrow}
\newcommand{\dn}{\downarrow}

\newcommand{\N}{\mathcal{N}}
\newcommand{\Tr}{\text{Tr}}

\newcommand{\bus}{{\tt BUS}}
\newcommand{\cinc}{{\tt CINC}}

\begin{document}

\title{Qudit Quantum Computation in the Jaynes-Cummings Model}


\author{Brian Mischuck} \email{brianm@phys.au.dk}
\author{Klaus M\o lmer}
\affiliation{Lundbeck Foundation Theoretical Center for Quantum System Research\\Department of Physics and Astronomy, Aarhus University}


\date{\today}

\begin{abstract}

We have developed methods for performing qudit quantum computation in the Jaynes-Cummings model with the qudits residing in a finite subspace of individual harmonic oscillator modes, resonantly coupled to  a spin-1/2 system. The first method determines analytical control sequences for the one- and two-qudit gates necessary for universal quantum computation by breaking down the desired unitary transformations into a series of state preparations implemented with the Law-Eberly scheme \cite{law_arbitrary_1996}.  The second method replaces some of the analytical pulse sequences with more rapid numerically optimized sequences. In our third approach, we directly optimize the evolution of the system, without making use of any analytic techniques. While limited to smaller dimensional qudits, the third approach finds pulse sequences which carry out the desired gates in a time which is much shorter than either of the other two approaches. 

\end{abstract}

\pacs{03.67.Bg, 03.67.Lx, 42.50.Dv,  85.25.Cp}
%\keywords{}

\maketitle

\section{\label{intro}Introduction}

The Jaynes-Cummings model \cite{strauch_arbitrary_2010, law_arbitrary_1996}, describing a harmonic oscillator coupled to a spin-1/2 system,  underlies a wide variety of potential platforms for quantum computation, such as atoms in cavities \cite{law_arbitrary_1996}, trapped ions \cite{cirac_quantum_1995,srensen_quantum_1999,childs_universal_2000, ben-kish_experimental_2003, kneer_preparation_1998}, superconducting circuits \cite{strauch_arbitrary_2010, strauch_entangled-state_2012, merkel_generation_2010, merkel_generation_2010, hofheinz_generation_2008, hofheinz_synthesizing_2009, wang_deterministic_2011, reed_high-fidelity_2010, mariantoni_implementingquantum_2011, strauch_quantum_2011, strauch_all-resonant_2012}, and clouds of cold atoms \cite{brion_quantum_2007}.   Because of its ubiquity, understanding how to control the Jaynes-Cummings model is a key step in the development of a quantum computer.
Synthesizing arbitrary states of one  \cite{law_arbitrary_1996, ben-kish_experimental_2003, kneer_preparation_1998, frana_santos_conditional_2001} or more \cite{strauch_arbitrary_2010,  strauch_entangled-state_2012,merkel_generation_2010} oscillators is a widely studied first step.  Growing interest in quantum computation, led to several proposals to use a harmonic oscillator as a bus between qubits \cite{cirac_quantum_1995, childs_universal_2000, srensen_quantum_1999}.  More recent work has focused on the controlability of the system  \cite{rangan_control_2004, yuan_controllability_2007}.  Here, the goal is to prove that arbitrary unitary transformations may be synthesized with a given set of controls, without necessarily providing an explicit algorithm to perform the synthesis.
The most advanced experimental implementation of the Jaynes-Cummings model is superconducting circuits, where both  state synthesis in single oscillators \cite{hofheinz_generation_2008, hofheinz_synthesizing_2009}, as well as entanglement between two oscillators \cite{merkel_generation_2010,wang_deterministic_2011} has been studied.  High fidelity qubit readout \cite{reed_high-fidelity_2010} as well as the quantum von Neumann architecture have also been demonstrated \cite{mariantoni_implementingquantum_2011}.

In the standard approach to quantum computing, information is stored in a series of two level qubits and the information is manipulated by applying one and two qubit gates.  In most schemes, single qubit gates can be done with relatively high fidelity,
but two qubit gates often cause problems both because control over two particle interactions is less well developed experimentally and because they can lead to increased coupling to the environment, leading to decoherence.  Thus, by minimizing multiparticle interactions, higher fidelity operations may be possible.
One approach to reducing the number of multiparticle interactions is to make use of $d$-level systems known as qudits.   By combining a few of the qubits into a single, larger dimensional system the gates between those qubits become manipulations of individual qudits, which may be more reliable.   A number of issues facing qudit quantum computation have already been considered, such as gate decompositions \cite{brennen_efficient_2006, brennen_criteria_2005, bullock_asymptotically_2005},  simulation \cite{nielsen_universal_2002},  and error correction \cite{grassl_efficient_2003}.   In addition, qudits may offer some advantages over qubits, in particular non-locality without entanglement \cite{bennett_quantum_1999},  improved detection efficiencies for Bell tests \cite{vrtesi_closingdetection_2010} and systems to study quantum chaos \cite{chaudhury_quantum_2009}.  A variety of experimental systems have been considered as qudits, including optical systems \cite{piani_linear-optics_2011, Lima:11}, superconductors \cite{strauch_all-resonant_2012, strauch_quantum_2011,  strauch_arbitrary_2010,  strauch_entangled-state_2012}, and atomic spins \cite{mischuck_control_2012,  merkel_constructing_2009,  merkel_quantum_2008, chaudhury_quantum_2007}.

ÒIn this paper we describe the synthesis of general unitary transformations on a qudit system defined as a two-level system and the first $n=0 \ldots N$ levels of a harmonic oscillator. Similar studies have been reported for the circuit QED system, where the interaction between the super conducting qubit system and the cavity field can only be switched off by detuning the systems with respect to each other \cite{strauch_quantum_2011,strauch_all-resonant_2012}. The Jaynes-Cummings model, however, also describes atomic systems coupled to a quantized cavity field  by a Raman process with a classical laser field that can be both detuned and switched completely off \cite{law_arbitrary_1996}. Harmonically trapped ions also implement the Jaynes-Cummings model with the possibility to resonantly drive an internal two-state transition, and a sideband transition, which couples the internal state and the motional oscillator state of the system \cite{ben-kish_experimental_2003}. Finally, the collective occupation of different internal states in an ensemble of atoms can be effectively described by oscillator degrees of freedom, and, e.g., by the  Rydberg blockade mechanism \cite{saffman_quantum_2010, brion_quantum_2007}, one of the populations may be effectively limited to two values and  thus implement a collective two-level degree of freedom in the system. These systems motivate our search for effective means to control the Jaynes-Cummings model, using the fast, resonant processes offered by the laser driven atomic systems. Our analysis uses a different approach and thus supplements recent work by Strauch \cite{strauch_all-resonant_2012}, which also includes use of resonant interactions.
As such, after introducing the basic controls in Sec.~\ref{basic_controls}, we show that resonant Jaynes-Cummings interactions are sufficient to generate arbitrary transformations on a qudit system in Sec.~\ref{analytic}.  We introduce a semi-analytic protocol to synthesize arbitrary transformations that uses a combination of numerical and analytic techniques to synthesize qudit transformation much more rapidly in Sec~\ref{Semi_analytic}.  We also show that direct numerical optimization, without making use of any analytic techniques, can speed up transformation synthesis even more in Sec.~\ref{fully_numerical_single_mode}.  Finally, in Sec.~\ref{two_mode} we extend our results to multiple modes.

\section{Basic controls \label{basic_controls}}

Our goal is to use qudits consisting of  the first $n=0,\ldots, N$ levels of a harmonic oscillator.  
% and perform arbitrary unitary transformations within the finite computational subspace, $h_{comp}$, consisting of the first $N+1$ states of the oscillators and the two spin states, represented in the following by product states $\{|n\uparrow\rangle, |n\downarrow\rangle\}_{n=0 \dots N}$.
To control the oscillators, we couple them to a spin-1/2 system, so that in the rotating frame and under the rotating wave approximation, the Hamiltonian of the Jaynes-Cummings model we consider is
\begin{align}
H = & H_s   + \sum_k H_{sc,k}, \label{controls}\\
H_s = & -\frac{1}{2}\Delta(t)\sigma_z + \frac{1}{2}\chi(t)\left(  \cos(\phi(t)) \sigma_x+\sin(\phi(t))\sigma_y \right),  \\
H_{sc,k} = & \frac{1}{2} g_k(t)(e^{i\beta_k(t)}a_k^\dagger \sigma_- +e^{-i\beta_k(t)} a_k\sigma_+).
\label{intro_controls}
\end{align}
The controls are the spin's detuning, $\Delta(t)$, drive strength and phase, $\chi(t),\ \phi(t)$, and coupling strength and phase, $g_k(t),\ \beta_k(t)$.   From this point forward, we will drop the explicit dependence on time in the controls.  These controls are available in systems such as clouds of Rydberg atoms  \cite{saffman_quantum_2010, brion_quantum_2007}, trapped ions \cite{kneer_preparation_1998, ben-kish_experimental_2003}  and three level atoms in cavities \cite{law_arbitrary_1996}.   Here, we focus on the generic features available in any system with these controls.   

We begin by considering the single mode case and drop the mode index $k$. It is convenient to define the finite computational subspace, $h_{comp}$, consisting of the first $N+1$ states of the oscillators and the two spin states, represented in the following by product states $\{|n\uparrow\rangle, |n\downarrow\rangle\}_{n=0 \dots N}$.   
The system's evolution can be greatly simplified in two regimes.  In the first regime $g = 0$, and the Hamiltonian 
\begin{equation}
H^{(1)} =-\frac{1}{2}\Delta\sigma_z  + \frac{1}{2}\chi\left(  \cos(\phi) \sigma_x+\sin(\phi)\sigma_y \right)
\end{equation}
only couples states within the two dimensional subspaces $h^{(1)}_n = \{\ket{n \up}, \ket{n \dn} \}$. Let $P^{(1)}_n = \ketbra{n}{n}$ denote the projectors onto those subspaces. Using $H^{(1)}$, we can generate an arbitrary rotation on the Bloch sphere,
\begin{equation}
\tilde{U}^{(1)}(\theta,\mathbf{n}) =  e^{-i\theta \, \mathbf{n}\cdot\boldsymbol{\sigma}/2}.
\label{U_1_plain}
\end{equation}

Next, we choose $\chi = 0$, so the Hamiltonian 
\begin{equation}
H^{(2)} = -\frac{1}{2}\Delta \sigma_z  +\frac{1}{2} g \left(e^{i\beta}a^\dagger\sigma_-  +e^{-i\beta}a\sigma_+  \right).
\end{equation}
couples the states within the two dimensional subspaces,
\begin{equation}
h^{(2)}_n = \left\{ \begin{array}{l l}
		 \{\ket{ n-1  \up}, \ket{n \dn} \}  & n\neq 0\\
		 \ket{0\dn} & n = 0.
	\end{array} \right.
\end{equation}
We also define the associated projectors,
\begin{equation}
P^{(2)}_n = \left\{ \begin{array}{l l}
		 \ketbra{n-1\up}{n-1\up}+\ketbra{n \dn}{n\dn}  & n\neq 0\\
		 \ketbra{0\dn}{0\dn} & n = 0
	\end{array} \right.
\end{equation}
and the set of Pauli operators on the $h^{(2)}_n$ subspaces with $n\neq0$,
\begin{align}
\sigma_{x,n}= &\ketbra{ n \dn}{ n-1 \up} + \ketbra{ n-1 \up}{ n \dn},\\
\sigma_{y,n}=& i \ketbra{ n \dn}{ n-1 \up} -i \ketbra{ n-1 \up}{n \dn},\\
\sigma_{z,n}= &- \ketbra{ n \dn}{n \dn} + \ketbra{n-1 \up}{n-1 \up}.
\end{align}
Defining $\sigma_{j,0}=0$, the Hamiltonian can be written
\begin{equation}
H^{(2)} = -  \frac{1}{2}\Delta\sigma_z+  \frac{1}{2}g\sum_{n=0}^\infty \sqrt{n}\left(\cos(\beta) \sigma_{x,n}  + \sin(\beta) \sigma_{y,n} \right),
\end{equation}
and it generates the following evolution,
\begin{equation}
 \tilde{U}^{(2)}(g,\Delta,\beta,T)  = e^{-i\left( -\Delta T\sigma_z + gT \sum\limits_{n=0}^\infty  \sqrt{n}  ( \cos(\beta)\sigma_{x, n}  +  \sin(\beta)\sigma_{y, n}  \right)    /2 }.
\label{U_2_plain}
\end{equation}
In the appendix, we show that it is possible to use these controls to synthesize transformations of the form
\begin{equation}
	U^{(2)} = \sum_{n=0}^{N+1} e^{-i \phi(n) \mathbf{m}(n)\cdot \boldsymbol{\sigma}_{n}/2   }P^{(2)}_n.
	\label{extension_2}
\end{equation}
Where $\phi(n)$ and $\mathbf{m}(n)$ are different rotation angles and torque vectors for each subspace $h^{(2)}_n$.  The transformations Eq.~(\ref{U_1_plain}), Eq.~(\ref{U_2_plain}) and Eq.~(\ref{extension_2}) form the basic building blocks from which all other controls will be built.

\section{Analytic synthesis of arbitrary transformations \label{analytic} }

Our unitary design scheme builds on the state preparation protocol originally developed by Law and Eberly \cite{law_arbitrary_1996}, which we review here for completeness.  Given the controls available, we need to synthesize a transformation, $U$, which maps an arbitrary state of the oscillator-spin system, $\ket{\phi}$, to $\ket{0\dn}$.
   To find $U$, we break the problem into a series of substeps,
\begin{equation}
U = \tilde{U}_{N+1}^{(1)}\prod_{k=1}^{N} \tilde{U}_k^{(2)} \tilde{U}_k^{(1)}.
\label{law_eberly}
\end{equation}
Where $\tilde{U}_j^{(1)}$ and $\tilde{U}_j^{(2)}$ have the form of  Eq.~(\ref{U_1_plain}) and (\ref{U_2_plain}).
The pulse sequences are chosen so that the population is removed sequentially from each harmonic oscillator level.  The state $\ket{\phi}$ initially has population spread over all $N$ oscillator levels,
\begin{equation}
\ket{\phi} = \sum_{n=0}^N c_{n\dn}\ket{n\dn} + c_{n\up}\ket{n\up}.
\end{equation}
The combination $\tilde{U}_1^{(2)} \tilde{U}_1^{(1)}$ is designed to remove all the population in $h^{(1)}_N$ by transferring the population to $\ket{N-1 \up}$.  First, $\tilde{U}_1^{(1)}$ transfers all the population in $h^{(1)}_N$ to $\ket{N \dn}$, then $\tilde{U}_1^{(2)}$ transfers all the population in $h^{(2)}_N$ to $\ket{N-1\up}$.  The result is a new state,
\begin{align}
\ket{\phi^{(1)}}= & \tilde{U}_1^{(2)} \tilde{U}_1^{(1)}\ket{\psi} \nonumber\\
= & \sum_{n=0}^{N-1} c_{n\dn}\ket{n\dn} + c_{n\up}\ket{n\up},
\end{align}
whose highest populated oscillator state is $N-1$.  The rest of the pulse sequences proceed in a similar manner with $\tilde{U}_j^{(2)} \tilde{U}_j^{(1)}$ clearing out each oscillator level one by one until all the population has been transferred to $h^{(1)}_0$.   A final $\tilde{U}^{(1)}_{N+1}$ transfers all of the population to $\ket{0\dn}$.

To see how the Law-Eberly scheme for state preparation can be extended to unitary transformation synthesis, we note that a unitary transformation may be defined as a transformation which maps a particular basis set $\ket{\phi_{n,s}}$ back to the $\ket{n,s}$ basis.  In other words, $U\ket{\phi_{n,s} }= \ket{n,s} $, where $n$ is the harmonic oscillator level and $s=\{\up,\dn \}$ is the spin state.

Thus, we can break a transformation up into a series of substeps,
\begin{equation}
U = \prod_n\prod_s U^{(ns)}.
\end{equation}
The main goal of each substep is to complete one state transformation, which maps a given state $\ket{\phi_{ns}   }$ back to the corresponding basis state $\ket{ns}$, so that
\begin{equation}
\ket{ns} =U^{(ns)}\ldots U^{(0\up)}U^{(0\dn)} \ket{\phi_{ns}   }.
\end{equation}
Each $U^{(ns)}$ accomplishes this goal via a Law-Eberly type sequence of substeps.  However, there are a couple of extra constraints on $U^{(ns)}$ which require some modifications of the Law-Eberly scheme.

After the first transformation, $U^{(0\dn)}$, we will have transformed $\ket{\phi_{0\dn}}$ to $\ket{0\dn}$.  All other $\ket{\phi_{ns}}$ will be transformed to a new set of target states $\ket{\phi_{ns}^{(1)}}  = U^{(0\dn)}\ket{\phi_{ns}}$.   We do not want population leaving the computational space, so $U^{(0\dn)}$ must satisfy the additional constraint that $\ket{\phi_{ns}^{(1)}}\in h_{comp}$.

The next pulse sequence, $U^{(0\up)}$, will transform $\ket{\phi_{0\up}^{(1)}}$ to $\ket{0\up}$, but we also need to ensure that the previously prepared state, $\ket{0\dn}$ is unchanged.  Thus, we also require that $U^{(0\up)}\ket{0\dn} = \ket{0\dn}$
For all other target states $U^{(0\up)}$ will transform them to $\ket{\phi_{ns}^{(2)}} = U^{(0\up)}\ket{\phi_{ns}^{(1)}}  $.  Once again, the transformation must prevent any population from leaking out of the computational space, so $\ket{\phi_{ns}^{(2)}}\in h_{comp}$.

All the other $U^{(ns)}$ will have a similar form.  First of all, $U^{(ns)}$ will complete the corresponding state preparation, so that
\begin{equation}
U^{(ns)} \ket{\phi^{(2n-s+1)}_{ns}} =  \ket{ns}.
\label{u_condition_0}
\end{equation}
Where we use $s=0$ for spin up and $s=1$ for spin down.
Secondly, we must keep track of the changes previous $U^{(ns)}$'s have made to the original target states.
The transformation $U^{(ns)}$  takes these states from $\ket{\phi^{(2n-s+1)}_{ns}}$ to some new set of target states, $\ket{\phi_{ns}^{(2n-s+2)}}$, which must remain in the computational subspace, in order to prevent population from leaking out of that space,
\begin{subequations}
\begin{align}
U^{(n\up)} \ket{\phi^{(2n-s+1)}_{n's'}} = &\ket{ \phi^{(2n-s+2)}_{n's'} }  \in h_{comp} \quad  n' > n \\
U^{(n\dn)} \ket{\phi^{(2n-s+1)}_{n's'}}  = & \ket{ \phi^{(2n-s+2)}_{ns} } \in h_{comp} \quad  n' > n \text{;  } n=n' \text{ and } s' = \up.
\end{align}
\label{u_condition_1}
\end{subequations}
Finally, after each $U^{(ns)}$, we will have a growing set of previously prepared states, $\{ \ket{0\dn},\ket{0\up},\ket{1\dn},\ldots \}$, and subsequent $U^{(ns)}$'s must not allow any further changes to these states,
\begin{subequations}
\begin{align}
U^{(n\dn)} \ket{n's'}  = & \ket{n's'}  \qquad\qquad\quad\; n' < n   \\
U^{(n\up)} \ket{ n's' } = & \ket{n's'} \qquad\qquad\quad\;   n' < n \text{;  } n=n' \text{ and } s'=\dn .
\end{align}
\label{u_condition_2}
\end{subequations}
We define $h_{id} = \{ \ket{0\dn},\ket{0\up},\ket{1\dn},\ldots \} $ to be those previously prepared states that must remain unchanged throughout the rest of the pulse sequence.
To fulfill Eq.~(\ref{u_condition_1})-(\ref{u_condition_2}) and carry out the requisite state preparation, all we need to do is break $U_{ns}$ up as we did in Eq.~(\ref{law_eberly})  

\begin{equation}
U^{(ns)} = \left\{ \begin{array}{ll}
	\prod_{k=1}^{N-n} U_k^{(ns,2)} U_k^{(ns,1)} & s = \up\\
	U_{N-n+1}^{(ns,1)}  \prod_{k=1}^{N-n} U_k^{(ns,2)} U_k^{(ns,1)} & s = \dn,
	\end{array} \right.
\end{equation}
where
\begin{subequations}
\begin{align}
U^{(ns,1)}_k = & \left\{ \begin{array}{ll}
	u^{(1)}_{N-k+1,n} & s = \up\\
	u^{(1)}_{N-k+1,n-1}  & s = \dn,
	\end{array} \right.\\
U^{(ns,2)}_k =& u^{(2)}_{N-k+1,n} ,
\end{align}
\label{U_nsa_in_terms_of_u}
\end{subequations}
and 
\begin{subequations}
\begin{align}
u^{(1)}_{j,\mathcal{N}} = & \sum_{l=0}^{\mathcal{N}} P^{(1)}_l+ M^{(1)}_j P^{(1)}_{j} + W^{(1)}_{comp}+W^{(1)}_{\perp comp} \label{first_u_1_j},\\
 u^{(2)}_{j,\mathcal{N}} = & \sum_{l=0}^{\mathcal{N}} P^{(2)}_l +M^{(2)}_j P^{(2)}_{j}  + W^{(2)}_{comp}+W^{(2)}_{\perp comp}\label{first_u_2_j}.
\end{align}
\label{first_u_j}
\end{subequations}
We show in the appendix how to synthesize $u^{(1)}_{j,\mathcal{N}}$  and $u^{(2)}_{j,\mathcal{N}}$ with the available controls. 

By inspection $U_k^{(ns,1)}$ and $U_k^{(ns,2)}$ above satisfy Eq.~(\ref{u_condition_1})-(\ref{u_condition_2}). They satisfy Eq.~(\ref{u_condition_0}) if $M^{(1)}_j$ and $M^{(2)}_j$ are chosen as described in the following, and if
the unitary transformations $W^{(2)}_{comp}$ and $W^{(1)}_{comp}$  operate within the subspaces of $h_{comp}$ orthogonal to $h_{0}^{(2)}\oplus h_{1}^{(2)}\oplus h_{2}^{(2)}\oplus \ldots \oplus h_{\mathcal{N}}^{(2)}\oplus h_{j}^{(2)}$ and $h_{0}^{(1)}\oplus h_{1}^{(1)}\oplus h_{2}^{(1)}\oplus \ldots \oplus h_{\mathcal{N}}^{(1)}\oplus h_{j}^{(1)}$, respectively, and the unitary transformations $W^{(2)}_{\perp comp}$ and $W^{(1)}_{\perp comp}$ operate only on the subspace orthogonal to $h_{comp}$.

\begin{figure}
\centering
 \subfloat[][]{ \includegraphics[width=0.75\textwidth]{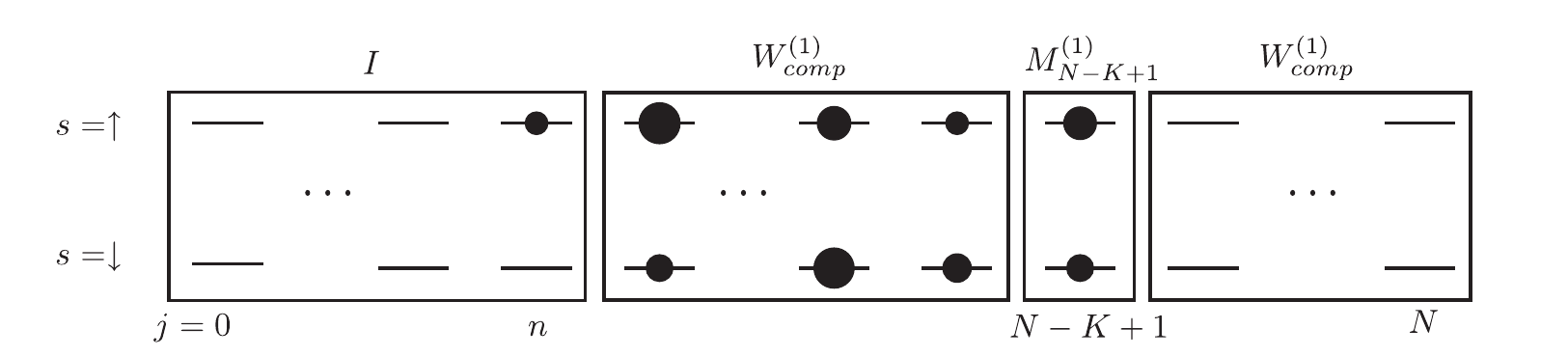}\label{U_ns1_K_s=up_step_beta}
 }\\
 \subfloat[][]{ \includegraphics[width=0.75\textwidth]{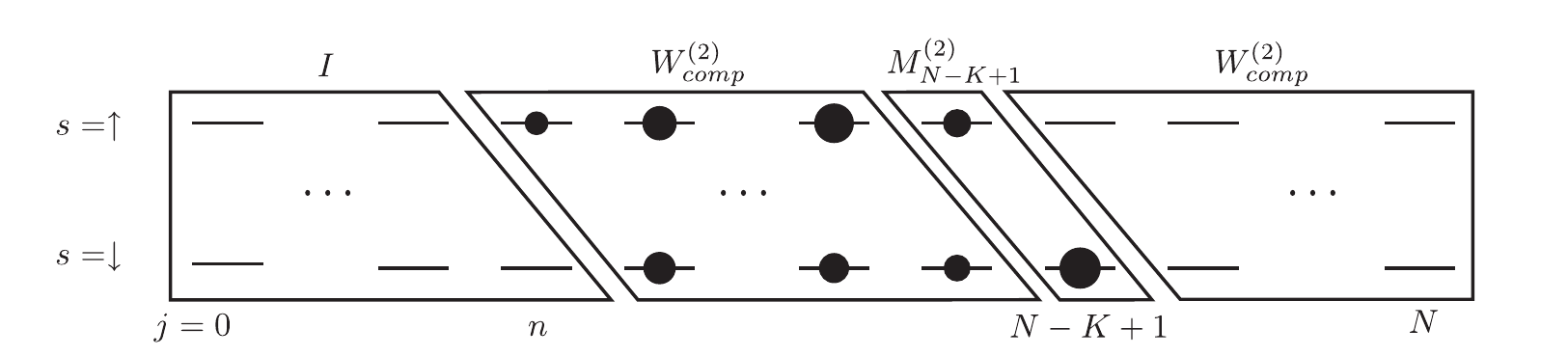}\label{U_ns2_K_s=up_step_alpha}
 }\\
  \subfloat[][]{ \includegraphics[width=0.75\textwidth]{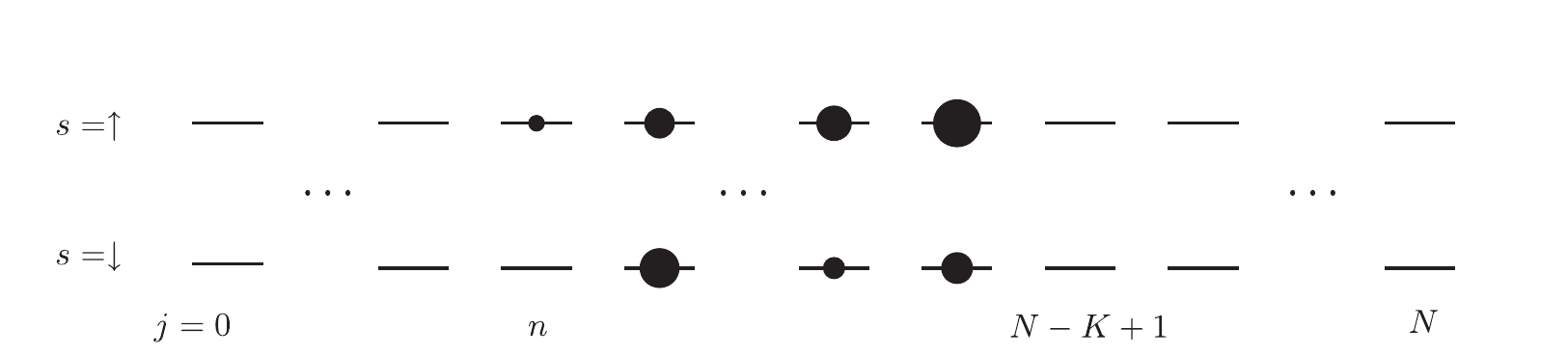}\label{U_ns2_K_s=up_step_beta}
 }
 \caption{   The effect of $U^{(ns,2)}_K U^{(ns,1)}_K$ on $\prod_{k=1}^{K-1} U^{(ns,2)}_k U^{(ns,1)}_k \ket{\phi^{(2n-2+1)}_{ns}}$.   The size of the circles represents the amount of population in each level. (a)  The transformation $U^{(ns,1)}_K$ leaves $h_{id}$ unchanged, while transferring all the population in $h^{(1)}_{N-K+1}$ of $\prod_{k=1}^{K-1} U^{(ns,2)}_k U^{(ns,1)}_k \ket{\phi^{(2n-2+1)}_{ns}}$ to $\ket{N-K+1\dn}$. (b)  The transformation $U^{(ns,2)}_K$ again leaves $h_{id}$ unchanged, while transferring all the population in $h^{(2)}_{n-K+1}$ of $U^{(ns,1)}_K\prod_{k=1}^{K-1} U^{(ns,2)}_k U^{(ns,1)}_k \ket{\phi^{(2n-2+1)}_{ns}}$ to $\ket{N-K \up}$. (c)  The net effect of $U^{(ns,2)}_K U^{(ns,1)}_K$ on $\prod_{k=1}^{K-1} U^{(ns,2)}_k U^{(ns,1)}_k \ket{\phi^{(2n-2+1)}_{ns}}$ is to remove all the population from the highest occupied oscillator level,  $N-K+1$, while leaving $h_{id}$ unaffected.  
 }
 \label{U_ns2_K_U_ns1_K}
\end{figure}

We begin by considering the $K^{th}$ step synthesizing $U^{(ns)}$ with $s=\up$,
as depicted in Fig.~(\ref{U_ns2_K_U_ns1_K}).  The population of  $\prod_{k=1}^{K-1} U_k^{(ns,2)} U_k^{(ns,1)} \ket{ \phi_{ns}^{(2n-s+1)} }$ is spread throughout the computational space, with the size of the circles representing the amount of population in each level.  By unitarity,   $\prod_{k=1}^{K-1} U_k^{(ns,2)} U_k^{(ns,1)} \ket{ \phi_{ns}^{(2n-s+1)} }$ has no overlap with the states in $h_{id}$.
Fig.~(\ref{U_ns1_K_s=up_step_beta}) depicts the action of $U^{(ns,1)}$.  It leaves $h_{id}$ unchanged while $M^{(1)}_{N-K+1}$ transfers all the population of  $\prod_{k=1}^{K-1} U_k^{(ns,2)} U_k^{(ns,1)} \ket{ \phi_{ns}^{(2n-s+1)} }$  within $h^{(1)}_{N-K+1}$ to $\ket{N-K+1 \dn}$.
Fig.~(\ref{U_ns2_K_s=up_step_alpha}) depicts the action of $U^{(ns,2)}_K$, which once again does not effect $h_{id}$, while   $M^{(2)}_{N-K+1}$ transfers all the population in $h^{(2)}_{N-K+1}$ of $U_K^{(ns,1)} \prod_{k=1}^{K-1} U_k^{(ns,2)} U_k^{(ns,1)} \ket{ \phi_{ns}^{(2n-s+1)} }$ to $\ket{N-K \up}$.  The net result of the $K^{th}$ step synthesizing $U^{(ns)}$ is shown in Fig.~(\ref{U_ns2_K_s=up_step_beta}).  As we see, all the population in $h^{(1)}_{N-K+1}$ of $\prod_{k=1}^{K-1} U_k^{(ns,2)} U_k^{(ns,1)} \ket{ \phi_{ns}^{(2n-s+1)} }$ has been transferred to $h^{(1)}_{N-K}$.
If $K\neq N-n$, then we are only concerned with transferring population, so the torque vectors of neither $M^{(1)}_{N-K+1}$ nor $M^{(2)}_{N-K+1}$ require a $z$-component.  In the final step of synthesizing $U^{(n\up)}$, the transformation $M^{(2)}_{n+1}$ must transfer all the population in $h^{(2)}_{n+1}$ of $U_{N-n}^{(ns,1)} \prod_{k=1}^{N-n-1} U_k^{(ns,2)} U_k^{(ns,1)} \ket{ \phi_{ns}^{(2n-s+1)} }$ to  $\ket{n \up}$ and set the resulting phase in $\ket{n \up}$ to zero. In this case, we must control the phase of the final state, so $M^{(2)}_{n+1}$  requires a torque vector with a $z$-component.

Similarly, If $s=\dn$ and $K\neq N-n +1$, then $M^{(2)}_{N-K+1} M^{(1)}_{N-K+1}$  must transfer all the population in $h^{(1)}_{N-K+1}$ of $\prod_{k=1}^{K-1} U_k^{(ns,2)} U_k^{(ns,1)} \ket{ \phi_{ns}^{(2n-s+1)} }$ to  $h^{(1)}_{N-K}$.  Since we are only concerned with transferring population, the torque vectors of neither $M^{(1)}_{N-K+1}$ nor $M^{(2)}_{N-K+1}$ require a $z$-component.
During the last step synthesizing $U^{(n\dn)}$,   the transformation $ M^{(1)}_{n}$ must move all the population in $h^{(1)}_{n}$ of $\prod_{k=1}^{N-n} U_k^{(ns,2)} U_k^{(ns,1)} \ket{ \phi_{ns}^{(2n-s+1)} }$ to  $\ket{n \dn}$ and set the resulting phase in $\ket{n \dn}$ to zero.  In this case, we must control the phase of the final state, so $M^{(1)}_{n}$  requires a torque vector with a $z$-component.

\section{Semi-analytic synthesis of arbitrary transformations   \label{Semi_analytic}   }
\subsection{Semi-analytic protocol   \label{Semi_analytic_construction}    }

The analytic construction of the previous section is sufficient to show that the coupled oscillator-spin system can be controlled in a finite time but we note that the number of operations needed to carry out  a general unitary operation on the qudit space is prohibitively large. In this section we will explore numerical optimization in order speed up the time required to synthesize an arbitrary transformation. 
In particular, fast implementations of the unitary operators $u^{(a=1,2)}_{j\N}$ in Eq.~(\ref{first_u_j}) can substantially shorten the amount of time required to perform general transformations.  For each target transformation $U$, we would, however, have to do a separate optimization to find the corresponding set of $u_{j,\N}^{(a)}$. We therefore proceed by synthesizing first the following transformations, which are independent of the chosen target transformation,
 \begin{subequations}
 \begin{align}
 V_{n,\N}^{(1)} = & \left\{ \begin{array}{l l}
 	 \sum_{j=0}^{\N} P^{(1)}_j+ e^{i\pi\sigma_z/2}P^{(1)}_n+ W^{'(1)}_{comp} + W^{'(1)}_{\perp comp} & n \neq \N+1\\
	 \sum_{j=0}^{\N} P^{(1)}_j+ e^{-i\pi/2}e^{i\pi\sigma_z/2}P^{(1)}_n + W^{'(1)}_{comp} + W^{'(1)}_{\perp comp} & n=\N+1, n\neq N\\
	 \sum_{j=0}^{\N} P^{(1)}_j+ e^{-i\pi/2}e^{i\pi\sigma_z/2}P^{(1)}_N  + W^{'(1)}_{\perp comp} & n=\N+1, n= N,
	 \end{array} \right. \label{V_1_n} \\
 V_{n,\N}^{(2)} = & \sum_{j=0}^{\N}  P^{(2)}_j+e^{-i\pi\sigma_{z,j}/2}P^{(2)}_{n} +P^{(2)}_{N+1}+ W'^{(2)}_{comp}+W'^{(2)}_{\perp comp}.    \hspace{90pt} \label{V_2_n}
 \end{align}
 \label{V_a_n}
\end{subequations}
Where $W^{'(a)}_{comp}$ and $W^{'(a)}_{\perp comp}$ are defined analogously to $W^{(a)}_{comp}$ and $W^{(a)}_{\perp comp}$ and their specific forms are not important for the rest of the construction.  
Thus, $V_{n,\N}^{(a)}$  is the identity within $h_{id}^{(a)}$, a $z$-rotation within $h_n^{(a)}$, and does not couple any population out of the computational space.
For the following constructions, we will only need $n\geq\N+1$.  All of the transformations $V_{n,\N}^{(a)}$ are special cases of Eq.~(\ref{extension_2}),  so we have already proven we can synthesize them with the available controls. 
Once we choose the dimension of our qudit, we run a single set of optimizations to find the required $V_{n,\N}^{(a)}$ for that dimension.

\begin{figure}
\centering
  \subfloat[][]{ \includegraphics{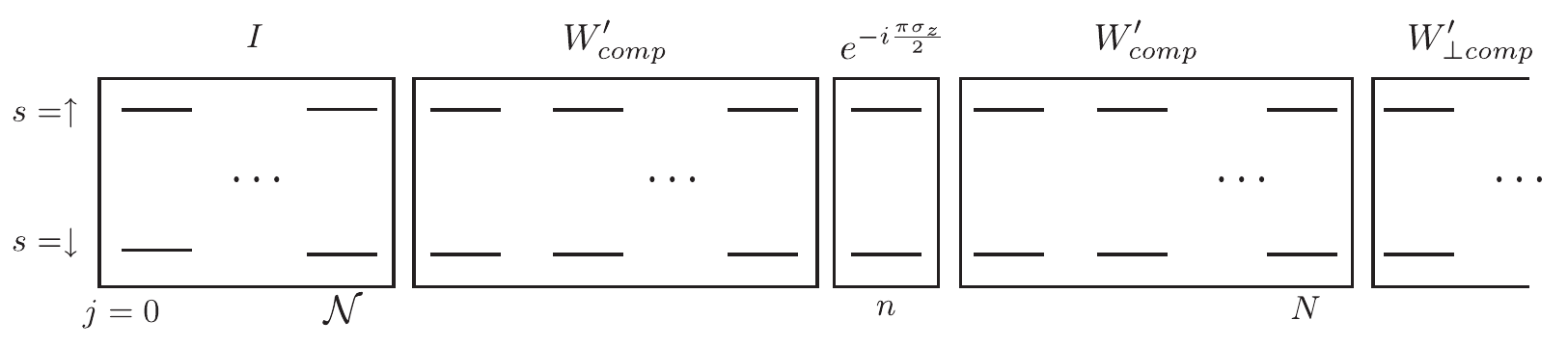}\label{V_1_n_sN_n_neq_sNplus1_n_neq_N}
 }\\
 \caption{ Illustration of $V^{(1)}_{j,\cal{N}}$ for $\cal{N} +$ $1 < n \leq N $ 
 }
 \label{V_1}
\end{figure}

We will need two slightly different types of pulse sequences to synthesize $U_k^{(ns,1)}$ and $U_k^{(ns,2)}$, depending on whether  the torque vector of  $M^{(1)}_{j}$ or $M^{(2)}_{j}$ requires a $z$-component. We begin with pulse sequences which use $ V_{n,\N}^{(1)}$ and $ V_{n,\N}^{(2)}$ to perform arbitrary rotations around a torque vector in the $xy$-plane of the $h_n^{(1)}$ and $h_n^{(2)}$ subspace, respectively, while also satisfying Eq.~(\ref{u_condition_1})-(\ref{u_condition_2})
\begin{subequations}
\begin{align}
 u_{n,\N}^{(1)} = & V_{n,\N}^{(1)\dagger} \sqrt{\tilde{U}^{(1)}}^\dagger V_{n,\N}^{(1)} \sqrt{\tilde{U}^{(1)}}\label{u_1_n_syn},\\
 u_{n,\N}^{(2)} = & V_{n,\N}^{(2)\dagger} \sqrt{\tilde{U}^{(2)}}^\dagger V_{n,\N}^{(2)} \sqrt{\tilde{U}^{(2)}}\label{u_2_n_syn}.
\end{align}
\label{u_n_syn_no_z}
\end{subequations}
To understand why Eq.~(\ref{u_n_syn_no_z}) holds, we begin by noting that if $\ket{ks} \in h_{id}^{(2)}$ then $ \sqrt{\tilde{U}^{(2)}} \ket{ks} \in h_{id}^{(2)}$, so $ V_{n,\N}^{(2)} \sqrt{\tilde{U}^{(2)}} \ket{ks} = \sqrt{\tilde{U}^{(2)}} \ket{ks}$, and it follows that
\begin{equation}
V_{n,\N}^{(2)\dagger} \sqrt{\tilde{U}^{(2)}}^\dagger V_{n,\N}^{(2)} \sqrt{\tilde{U}^{(2)}} \ket{ks} =  \ket{ks}.
\end{equation}
Similarly, $V_{n,\N}^{(2)} \sqrt{\tilde{U}^{(2)}} \ket{ns} = e^{-i\pi\sigma_{z,n}/2} \sqrt{\tilde{U}^{(2)}} \ket{ns}$ so that
\begin{equation}
V_{n,\N}^{(2)\dagger} \sqrt{\tilde{U}^{(2)}}^\dagger V_{n,\N}^{(2)}\sqrt{ \tilde{U}^{(2)}} \ket{ns} = \tilde{U}^{(2)} \ket{ns}.
\end{equation}
Neither $V_{n,\N}^{(2)}$ nor $ \sqrt{\tilde{U}^{(2)}}$ couple states in the computational subspace to states outside of it, so for $\ket{\phi}\in h_{comp}$,
\begin{equation}
V_{n,\N}^{(2)\dagger} \sqrt{\tilde{U}^{(2)}}^\dagger V_{n,\N}^{(2)}\sqrt{ \tilde{U}^{(2)}} \ket{\phi} \in h_{comp}.
\end{equation}
Finally, with $ M^{(2)}_n =\tilde{U}^{(2)}$, we obtain Eq.~(\ref{u_1_n_syn}), and a similar argument yields Eq.~(\ref{u_2_n_syn}).

The pulse sequence which synthesizes torque vectors with $z$-components within $h_n^{(1)}$ and $h_n^{(2)}$ is a modification of the pulse sequence just discussed,
 \begin{subequations}
  \begin{align}
  u_{n,\N}^{(1)} = & \tilde{U}^{(1)'\dagger} V_{n,\N}^{(1)\dagger} \sqrt{\tilde{U}^{(1)}}^\dagger V_{n,\N}^{(1)} \sqrt{\tilde{U}^{(1)}}\tilde{U}^{(1)'} \label{u_1_n_syn_z},\\
 u_{n,\N}^{(2)} = & \tilde{U}^{(2)'\dagger} V_{n,\N}^{(2)\dagger} \sqrt{\tilde{U}^{(2)}}^\dagger V_{n,\N}^{(2)} \sqrt{\tilde{U}^{(2)}} \tilde{U}^{(2)'}\label{u_2_n_syn_z}.
 \end{align}
 \label{u_n_syn_z}
 \end{subequations}

If we choose
\begin{align}
\tilde{U}^{(1)} =& e^{-i \alpha\left( \cos(\nu)\sigma_x   + \sin(\nu)\sigma_y     \right)/2}, \\
\tilde{U}^{(1)'} =&e^{i \eta \sigma_x /2},
\end{align}
then $M_n^{(1)} = \tilde{U}^{(1)\dagger'} \tilde{U}^{(1)} \tilde{U}^{(1)'}  $ has the required $z$-component.  Similar choices for $ \tilde{U}^{(2)}$ and $\tilde{U}^{(2)'}$ will give a $z$-component to $M^{(2)}_n =  \tilde{U}^{(2)'\dagger} \tilde{U}^{(2)} \tilde{U}^{(2)'}  $.    
Finally, we note that since $\sqrt{\tilde{U}^{(1)}}\tilde{U}^{(1)'} =  \tilde{U}^{(1)''} $ is of the form Eq.~(\ref{U_1_plain}), we use the following slightly shorter sequence,
 \begin{equation}
  u_{n,\N}^{(1)} =  \tilde{U}^{(1)'\dagger} V_{n,\N}^{(1)\dagger} \sqrt{\tilde{U}^{(1)}}^\dagger V_{n,\N}^{(1)} \tilde{U}^{(1)''} .
 \label{u_1_compress}
 \end{equation}
Thus, Eq.~(\ref{u_n_syn_no_z}), Eq.~(\ref{u_2_n_syn_z}) and Eq.~(\ref{u_1_compress}) are sufficient to synthesize arbitrary unitary transformations.

\subsection{Numerical optimization \label{num_opt_sa}}

Our goal in this section is to find a pulse sequence, $V^{(a)}_{n\N,sa}$, which approximates $V^{(a)}_{n\N}$ using fewer pulses than the analytic sequence described in Sec.~\ref{analytic}.  To do so, we numerically optimize pulse sequences of the form
\begin{align}
V^{(a)}_{n,\N,sa} = & \prod_{m=1}^M \tilde{U}^{(2)}(g^{(a)}_{n,\N,m},0,\beta^{(a)}_{n,\N,m},T_g/2).
\label{sa_pulse_sequence}
\end{align}
Where $T_g = 2\pi/g_{max}$ and $g_{max}$ is the maximum allowed coupling strength.  The optimization finds a sequence of $\{g^{(a)}_{n,\N,m},\beta^{(a)}_{n,\N,m}\}$ which minimizes the error up to a global phase change
\begin{equation}
\epsilon^{(a)} = \frac{1}{4(N+1)}\min_\phi \left| \left| P_C\left( V^{(a)}_{n,\N} - e^{i\phi} V^{(a)}_{n,\N,sa} \right)P_C \right| \right|^2.
\end{equation}
$V^{(a)}_{n,\N}$ is only defined on a subspace of the computational space, and we can hence ignore the unimportant subspace.  We begin by noting that because $\Tr\{P_C V^{(a)\dagger}_{n\N,sa} P_C V^{(a)}_{n\N,sa} P_C\} \leq  2(N+1)$, we have
\begin{equation}
 \epsilon^{(a)} \leq 1-\frac{1}{2(N+1)} | \Tr\{ P_C V^{(a)\dagger}_{n\N} P_C V^{(a)}_{n\N,sa} P_C \}|.
\end{equation}
If we define $P^{(a)}_{n\N}$ to be the projector onto the optimized subspace, then
\begin{subequations}
\begin{empheq}[left ={  P_{n,\N}^{(a)} =  \empheqlbrace}]{alignat=2}
&\sum_{j=1}^{\N+1} P^{(2)}_j+P^{(2)}_{n} +P^{(2)}_{n+1} + P^{(2)}_{N+1} & \quad a = 1, \,n\neq \N+1,\,n \neq N \\
& \sum_{j=1}^{\N+1} P^{(2)}_{j} +P^{(2)}_{N} + P^{(2)}_{N+1} &\quad a = 1, n \neq \N+1,\, n = N  \\
& \sum_{j=1}^{\N} P^{(2)}_{j}+P^{(2)}_{n} + P^{(2)}_{n+1} + P^{(2)}_{N+1}  &\quad a = 1,  n=\N+1,\,n \neq N  \\
& \sum_{j=1}^{\N} P^{(2)}_{j}+P^{(2)}_{N} + P^{(2)}_{N+1}    &\quad a = 1, n=\N+1,\, n = N \\
& \sum_{j=1}^{\N}  P^{(2)}_j+P^{(2)}_{n} +P^{(2)}_{N+1}  &\quad a = 2,  \hspace{101pt}
\end{empheq}
\label{proj_opt_space}
\end{subequations}
 and we arrive at the following error to be minimized
\begin{equation}
\epsilon^{(a)}_{sa} = 1-\frac{1}{2(N+1)}|d^{(a)}_{\perp n\N} + \Tr\{P^{(a)}_{n\N} V^{(a)\dagger}_{n\N} P^{(a)}_{n\N} V^{(a)}_{n\N,sa} P^{(a)}_{n\N}|.
\end{equation}
Where $d^{(a)}_{\perp n\N} $ is the size of the subspace of the computational space orthogonal to $P^{(a)}_{n\N}$.
Note that the sums in Eq.~(\ref{proj_opt_space}) begin with $j=1$ while the sums in Eq.~(\ref{V_a_n}) begin with $j=0$.  The reason is that the form of $\tilde{U}^{(2)}$ ensures that any sequence of the form Eq.~(\ref{sa_pulse_sequence}) will leave $\ket{0\dn}$ unchanged.  Thus, we do not need to explicitly account for $\ket{0\dn}$ when we perform the optimizations.

For each $V^{(a)}_{n\N,sa}$, we begin with $M=3$ pulses and perform at most 40 optimizations using MATLAB's constrained optimization routine, $fmincon$,  with the constraint that $g^{(a)}_{n,m} \leq g_{max}$.
Each optimization finds a choice of $\{g^{(a)}_{n,m}$, $\beta^{(a)}_{n,m}\}$ that is a local minimum of $\epsilon^{(2)}_{sa}$.
We repeat this entire procedure, incrementing $M$ by one each time until we find a pulse sequence which approximates $V^{(a)}_{n\N}$ to an error better than a chosen threshold, $\epsilon_{threshold}$.
Since our end goal is to find a sequence of pulses which approximates our target transformation to a desired accuracy, $\eta$,
and we need to apply $g_{sa}(N) = 4N^2+6N+2$ transformations of the form $V^{(a)}_{n\N}$ to use the construction of Sec.~\ref{Semi_analytic_construction}, our threshold is
\begin{equation}
\epsilon_{threshold} = \frac{\eta}{g_{sa}^{2}(N)}.
\label{threshold_error}
\end{equation}
Finally, to bound the total time to synthesize an arbitrary transformation, we assume $g_{max} = \chi_{max}$, so we can bound the time for $\tilde{U}^{(a)}$ by $T(\tilde{U}^{(a)})~\leq ~\frac{1}{2} T_g$ and $T\left(\sqrt{\tilde{U}^{(a)}   }\right)\leq \frac{1}{2} T_g$ in Eq.~(\ref{u_n_syn_no_z}) and Eq.~(\ref{u_n_syn_z}).

\begin{figure}[!ht]
\centering
 \subfloat[][]{ \includegraphics[width=0.5\textwidth]{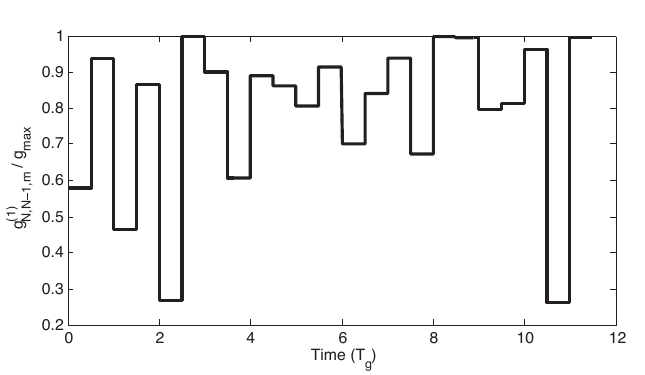}\label{theta_sa}
 }
 \subfloat[][]{ \includegraphics[width=0.5\textwidth]{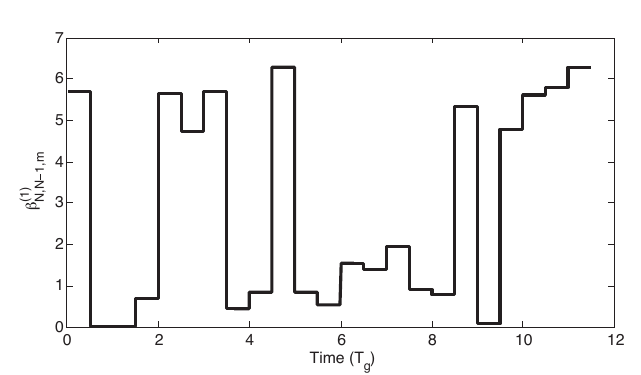}\label{phi_sa}
 }
 \caption{Controls for $V^{(1)}_{N,N-1}$ for $N=8$.  Fig.~(\ref{theta_sa}) depicts the coupling strength per pulse, $g^{(1)}_{N,N-1,m}$, while Fig.~(\ref{phi_sa}) depicts the angle of the torque vector relative to the $x$-axis of the Bloch sphere, $\beta^{(1)}_{N,N-1,m}$.   The optimization finds a sequence of $\{g^{(a)}_{n\N,m},\beta^{(a)}_{n\N,m}\}$  as in Eq.~(\ref{sa_pulse_sequence}) that minimizes the resulting error, $\left| \left| V^{(a)}_{n\N} - V^{(a)}_{n\N,sa} \right| \right| $.   The optimization is repeated a number of times with initial random guesses for $\{g^{(a)}_{n\N,m},\beta^{(a)}_{n\N,m}\}$ and increasingly longer pulses, until a threshold error $\eta \leq 10^{-4}$, as defined in Eq.~(\ref{threshold_error}), is achieved.
  }
 \label{sa_controls}
\end{figure}

In Fig.~(\ref{sa_controls}) we plot the controls found by numerical optimization to synthesize $V^{(1)}_{N,N-1}$ for $N=8$.  As is typical for numerically optimized controls, there is not much structure in $g^{(1)}_{N,N-1,m}$.  However, examination of $\beta^{(1)}_{N,N-1,m}$ reveals that the torque vector lies near the $x$-axis for much of the synthesized transformation.

\begin{figure}
\includegraphics[width=0.5\textwidth]{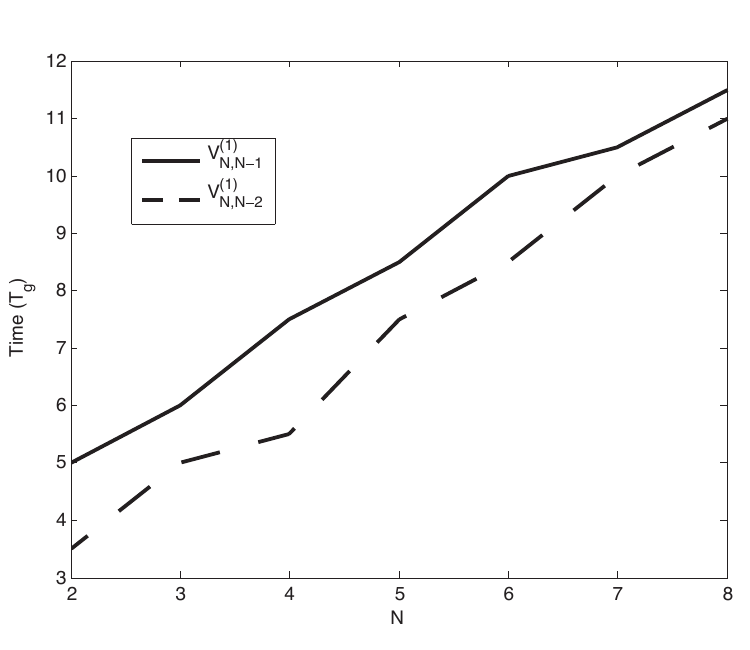}
\caption{ Time, in units of $T_g$, required to reach a threshold of $\eta \leq 10^{-4}$ for two families of transformations, $V^{(1)}_{N,N-1}$ and $V^{(1)}_{N,N-2}$, where $N$ is the highest controlled oscillator level}
\label{two_V_1}
\end{figure}

\begin{figure}[h]
\centering
 \subfloat[][]{ \includegraphics[width=0.5\textwidth]{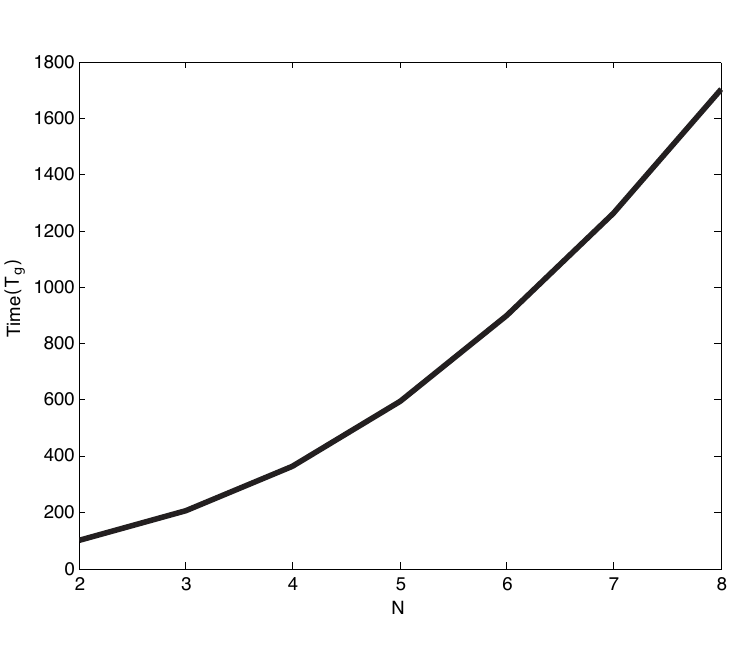}  \label{sa_time}
 }
 \caption{Time, in units of $T_g$, to synthesize an arbitrary target transformation, $U$, using the semi-analytic construction versus the highest controlled oscillator level, $N$.  The semi-analytic pulse sequence replaces the most time consuming pulses of the analytic construction with the numerically optimized pulses $V^{(a)}_{n\N,sa}$.
 The time plotted is the minimum time found to approximate $U$ to the target error of $\eta \leq10^{-4}$. 
 }
 \label{sa_transformation}
\end{figure}

Results from the optimizations are presented in Fig.~(\ref{two_V_1}) and Fig.~(\ref{sa_transformation}).  In Fig.~(\ref{two_V_1}) we plot the time required to reach a threshold of $\eta \leq 10^{-4}$ for two families of transformations, $V^{(1)}_{N,N-1}$ and $V^{(1)}_{N,N-2}$ used to construct more arbitrary transformations.  We are able to reach high fidelity transformations in a time less than $12T_g$ in all cases. 
 In Fig.~(\ref{sa_time}) we plot the time $T(U_{sa})$ required by the semi-analytic construction in units of $T_g$ for a target error of $\eta \leq10^{-4}$.
Although a large number of pulses is required for the semi-analytic construction, it is many orders of magnitude faster than the fully analytical construction.

Since we are able to synthesize arbitrary unitaries in dimensions up to $d=2(N+1)$, we can compare our results to the time required to synthesize the same transformations using $n$ qubits.  For concreteness, we consider synthesizing an arbitrary transformation in a $d=16$ dimensional space consisting of either the $N=7$ case of our oscillator and spin system or four conventional qubits.

The best known decomposition of an arbitrary n-qubit transformations is given in \cite{shende_synthesis_2006}, and for 4 qubits, it requires a total of one hundred $\tt{CNOT}$s.
However, this result assumes any two qubits may be directly coupled while often two qubits can only be coupled via a bus qubit or oscillator degree of freedom. Thus for any $\tt{CNOT}$, two extra $\tt{SWAP}$ operations are needed so the total number of 2-qubit operations is closer to three hundred. The number of two-qubit operations is a lower bound on the total time, and if each 2-qubit gate requires a time on the order of $T_g$, we estimate the total time to implement an arbitrary 4-qubit gate with the best known gate decomposition to be bounded by $T_{gate} > 300T_g$.   Taking this estimate into account, we see from Fig.~(\ref{sa_time}) that when $N=7$, the semi-analytic approach of this section is only longer by a small factor than the bound on the multiqubit approach. A detailed comparison of the multiqubit and semi-analytic approach including all $\tt{SWAP}$s, single qubit gates, the optimum gate decomposition in our architecture and the role of decoherence is beyond the present scope of this article. 
However, we can already conclude that the performance of our semi-analytic approach makes qudit computing comparable with qubit implementations.

\section{Fully numerical synthesis of $\cinc'$ \label{fully_numerical_single_mode}}

While the preceding section shows that arbitrary unitaries can be synthesized more rapidly with the semi-analytic protocol, the time required may not be optimal.  To  find a pulse sequences that performs even better, we now investigate a fully numerical optimization where the coupling between the spin and the oscillator is held constant while we allow the spin detuning, $\Delta(t)$, Rabi frequency, $\chi(t)$, and phase, $\phi(t)$, to be stepwise constant during intervals of duration $dt$, for a total time, $T_f$.  The values of  $\Delta(t)$, $\chi(t)$ and $\phi(t)$ during each interval are optimized with MATLAB's $fminunc$, which performs an unconstrained optimization of $\chi(t)$, $\Delta(t)$ and $\phi(t)$.  In principle, this could yield values of the Rabi frequency and detuning which are not experimentally realistic.  However, we seed the initial random guess with $\chi(t) , \Delta(t) \leq 0.9 g$ and we find optimal controls with $\chi(t), \Delta(t) \le 2 g$. Unconstrained optimization does not result in $\chi(t),\Delta(t) \gg g$ because that would effectively decouple the spin and the oscillator, making it impossible to synthesize general unitary transformations.  For each $dt$ and $T_f$, 20  optimizations are performed, and the pulse sequence with the highest fidelity is chosen.  As a test case, the gate we have optimized is

\begin{equation}
{\tt CINC}' = I_{osc} \otimes \ketbra{\dn}{\dn} + \sum_{n=0}^N \ketbra{n\oplus1}{n}\otimes\ketbra{\up}{\up}.
\end{equation}
This gate increases the oscillator level by 1, modulo N+1, conditional on the spin being $s=\up$.  As we will see in Sec.~\ref{two_mode}, this gate can be used to construct two qudit gates.  The fidelity of the optimized gate, $U_{fn}$,  with the target gate,~$U$,  is
\begin{equation}
F = \frac{1}{[2(N+1)]^2}|Tr\{U_{fn}^\dagger U\}|^2 .
\end{equation}
For small errors, the gate fidelity can be related to the error defined above by
\begin{equation}
F = 1 -2\eta
\end{equation}

The optimizations have to be performed within a finite subspace and to ensure we can perform the optimizations in a reasonable amount of time, we keep this subspace as small as possible, and we penalize the leakage of population out of the computational space. In practice, we divide the simulated subspace into three regions given by $N<N_{pad}<N_{opt}$.  The highest oscillator level in the computational space is $N$ and the largest oscillator level we use in the optimization is $N_{opt}$.
We pad the calculation with a subspace lying above $N$ but below $N_{pad}+1$ and any population that leaks into this subspace is not penalized.   On the other hand, we penalize any population that leaks into the levels lying above $N_{pad}$  during the course of the evolution.  We can calculate the population which leaks out of $N_{pad}$ during the course of the evolution and average over all states in the computational space, $\ket{\psi}$.
\begin{equation}
L = \sum_j^P dt\, \int d\psi\, \left| \left| P_L U_{fn}(t_j) \ket{\psi}\right| \right|^2.
\end{equation}
Where $P$ is the total number of pulses, $U_{fn}(t_j)$ is the total evolution after the $j^{th}$ subpulse and $P_{L}$ is the projector onto the subspace which we wish to penalize,
\begin{equation}
P_L = \sum_{N_{pad}+1}^{N_{opt}}\ketbra{n}{n}.
\end{equation}
The integral over $\ket{\psi}$ can be simplified \cite{pedersen_fidelity_2007}, so the leakage is
\begin{equation}
L = \sum_j^P \frac{1}{2(N+1)(2N+3)}\left(\Tr\{M(t_j)M(t_j)^\dagger\} +|\Tr\{M(t_j)\}|^2 \right).
\end{equation}
Where
\begin{equation}
M(t_j) = P_C U_{fn}(t_j)^\dagger P_L U_{fn}(t_j) P_C.
\end{equation}
Then the optimized quantity is
\begin{equation}
C_{FN} = 1-F + w L.
\end{equation}
After finding a pulse sequence which minimizes $C_{FN}$ we calculate the system's evolution in a larger subspace, $N_{check}>N_{opt}$ and recalculate the gate fidelity to ensure that using only a finite number of oscillator levels does not effect the calculation.  We have found empirically that $w=100$ ensures that the leakage out of the computational space is sufficiently small that when we recalculate the fidelity in the lager space, it is still high enough.

\begin{figure}[!ht]
\centering
 \subfloat[][]{ \includegraphics{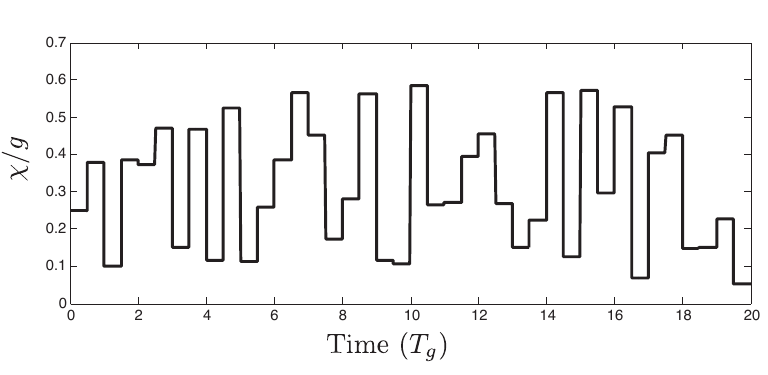}\label{CINCp_chi}
 }
 \subfloat[][]{ \includegraphics{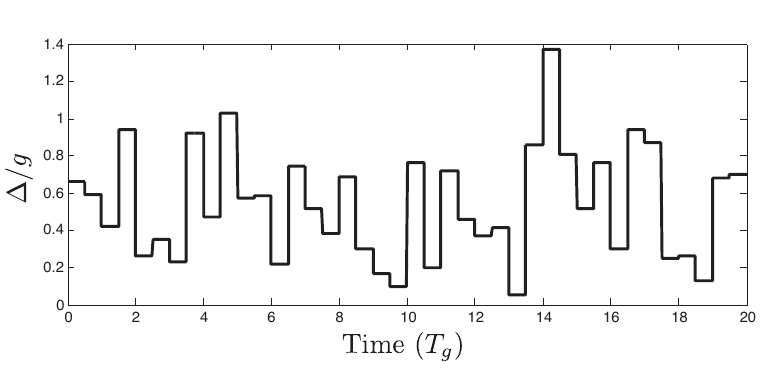}\label{CINCp_delta}
 }\\
 \subfloat[][]{ \includegraphics{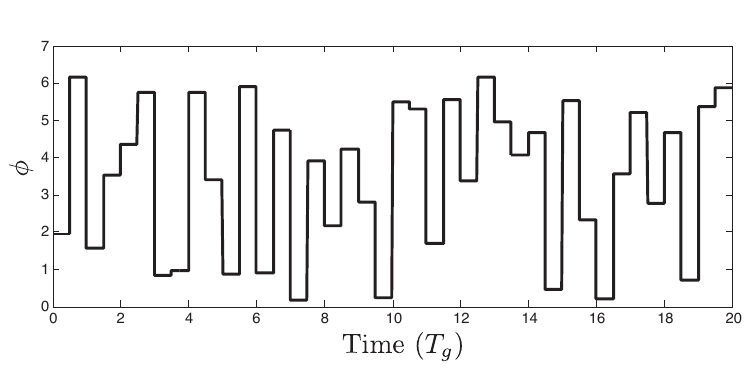}\label{aaaaaaaaa}
 }\\
 \caption{ Numerically optimized control pulses used to synthesize $\tt{CINC}'$ for the  $N=2$, $dt = 0.5T_g$, and $T_f = 20T_g$ case.
 }
 \label{CINCp_controls}
\end{figure}

In Fig.~(\ref{CINCp_controls}) we plot the controls for the $N=2$, $dt = 0.5T_g$, and $T_f = 20T_g$ case.  Although we perform an unconstrained optimization, we see in Fig.~(\ref{CINCp_chi}) and (\ref{CINCp_delta}) that neither $\chi$ nor $\Delta$ require values that are large compared to $g$, as anticipated.  Beyond that, there is no real structure to the controls, as is typical when they have been numerically optimized.

In Fig.~(\ref{CINCprime}) we plot the fidelity, $F$ versus time for $N=2,3$, with time in units of the vacuum Rabi period, $T_g$.  We perform the optimizations for two different sub-pulse lengths, $dt = 0.5T_g$ and $dt=T_g$.   We use $N_{pad} = N+3$, $N_{opt} = N+5$ and $N_{check} = 4N_{opt}$.  With $dt = 0.5 T_g$ we achieve errors on the order of $10^{-4}$ with tens of pulses for both $N=2$ and $N=3$, substantially shorter than the semi-analytic approach, cf. Fig.~(\ref{sa_transformation}).

\section{Two mode control \label{two_mode}}

In this section, we extend our results to cover arbitrary quantum gates between qudits stored in the modes of different harmonic oscillators.  We consider the system described by Eq.~(\ref{controls}), so that each oscillator is coupled to a central spin.  In the this case, it is more convenient to take the computational space to consist of the first $n=0 \ldots N$ levels of each oscillator.  Then if there are $m$ oscillators, the size of the quantum comptuer's Hilbert space is $(N+1)^m$.
Our controls are once again the time dependent $\chi$, $\phi$, $\Delta$, $g_k$ and $\beta_k$.  For simplicity, we assume only a single $g_k \neq 0$ at a time, so that only one oscillator is coupled to the central spin, and we optimize $\chi$, $\phi$ and $\Delta$.  Although the oscillators are never directly coupled, we can use the central spin as a bus to synthesize multiqudit gates, using the gates developed in previous sections.

In analogy to qubits, multiqudit gates can be synthesized from a universal gate set.   Along with arbitrary single qudit gates, the two-qudit $\cinc$ gate is sufficient to synthesize arbitrary multiqudit gates \cite{brennen_efficient_2006}.   Where $\cinc$ is defined as
\begin{equation}
{\tt CINC} = \sum_{n_1=0}^{N-1}\ketbra{n_1}{n_1} \otimes I_2 + \ketbra{N}{N} \otimes \sum_{n_2=0}^N \ketbra{n_2\oplus1}{n_2}.
\end{equation}
The action of ${\tt CINC}$ is similar to that of ${\tt CINC}'$, with the target qudit's level increased by one modulo $N$ if the control is in $\ket{N}$.  Because we have already shown we can synthesize arbitrary single qudit gates, showing that we can synthesize the two-qudit $\cinc$ is sufficient to prove we can synthesize arbitrary multiqudit gates.

In fact, we can synthesize ${\tt CINC}$ by using the spin as a bus between the two oscillators.  We first define,
\begin{equation}
{\tt BUS} = \sum_{n=0}^{N-1} P^{(2)}_n + \sigma_{x,N} + W.
\end{equation}
Where $W$ acts on the $h^{(2)}_{N+1}\oplus h^{(2)}_{N+2}\oplus \ldots$  subspace.   Thus, assuming the central spin is initialized to spin down, the two-qudit $\tt{CINC}$   can be synthesized from
\begin{equation}
\bus^\dagger_{s2} \cinc'_{1s} \bus_{s2}  = \cinc_{12} \otimes \ketbra{\dn}{\dn} + W_\up \otimes \ketbra{\up}{\up}.
\label{cinc_decomp}
\end{equation}
Where the form of $W_\up$ is unimportant.  In the above notation ${\tt M}_{ab}$ is meant to indicate that the target is system $a$ and the control is $b$.  Where $\{a,b\} = \{1,2,s\}$ is the first qudit, second qudit, and central spin.  Thus, $\bus_{s2}$ couples the spin and the second oscillator, while $ \cinc'_{1s}$ couples the spin and the first oscillator.  To ensure that only one oscillator couples to the spin we assume that $g_1=0$ while we synthesize $\bus_{s2}$ and  $g_2=0$ while we synthesize $\cinc'_{1s}$

We have seen in the previous section how to synthesize $\cinc'_{1s}$, which acts between the first qudit and the spin.  Since $\bus_{s2}$ is a particular $U^{(2)}$, we can synthesize it with the following sequence
\begin{equation}
\bus_{s2} \approx \prod_{m=1}^M \tilde{U}^{(2)}(g_{max},\Delta_m,0,dt).
\end{equation}
Where $\tilde{U}^{(2)}$ is taken to operate between the spin and second oscillator.
As in Sec.~\ref{fully_numerical_single_mode}, we choose to work with a constant pulse width $dt$ and numerically optimize $\Delta_m$ using the same procedure used to synthesize $V^{(a)}_{n,\N,sa}$ in Sec.~\ref{num_opt_sa}.  For each $N$ and $dt$, after the optimization   we choose the pulse sequence which achieves an error below $\epsilon^{(2)} \leq 10^{-4}$ with the fewest number of pulses.
Finally, we synthesize $\bus_{s2}^\dagger$ with
\begin{equation}
\bus_{s2}^\dagger \approx  \prod_{m=M}^1 \tilde{U}^{(2)}(g_{max},-\Delta_m,\pi,dt).
\end{equation}
Where the order of the product has been reversed, so that the $M^{th}$ pulse is applied first, and we use the same detunings, $\Delta_m$ used to synthesize $\tt{BUS}$.  Since $\Delta_m$ is one of our controls, we are free to reverse it's sign, and $\beta = \pi $ effectively reverses the sign of the coupling between the oscillator and the spin.

\begin{figure}[!ht]
\centering
 \subfloat[][]{\includegraphics[width=0.5\textwidth]{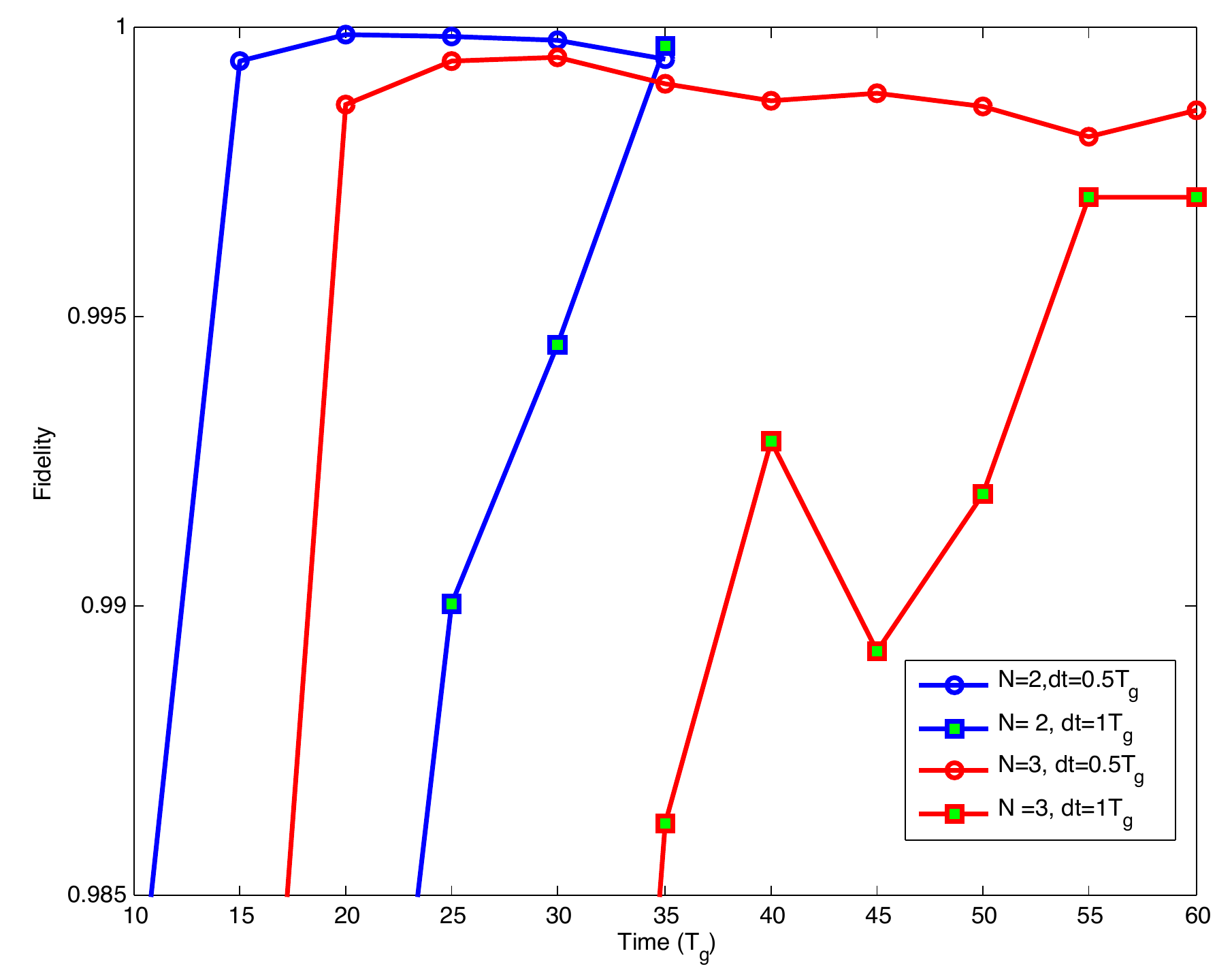}\label{CINCprime}
 }
 \subfloat[][]{\includegraphics[width=0.5\textwidth]{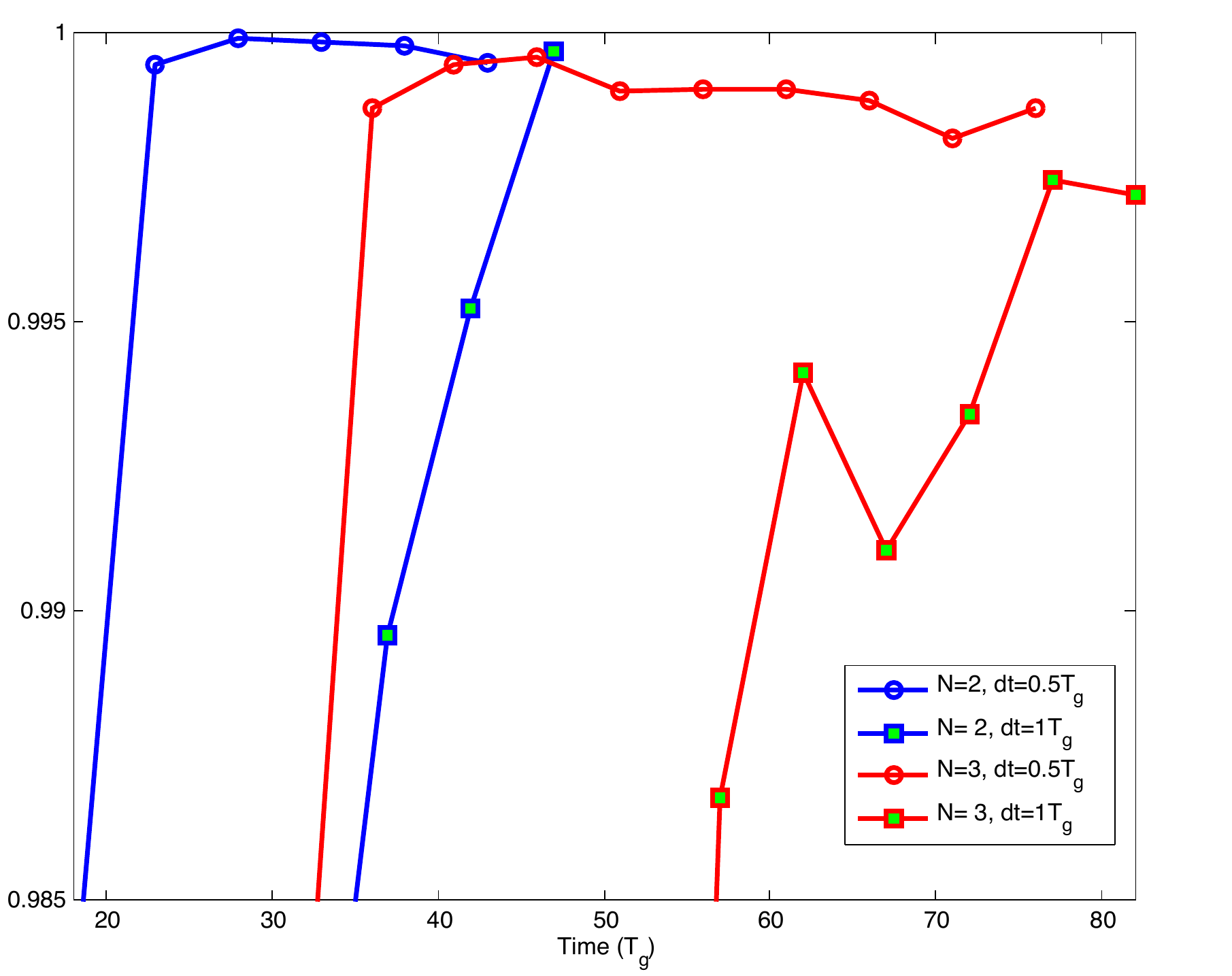}\label{CINC}
 }\\
 \caption{  Fidelity vs. time in units of $T_g$ for $N=2,3$.  In Fig.~(\ref{CINCprime}), we plot  the fidelity of ${\tt CINC}'$, which couples a single oscillator to the spin. The coupling strength between the spin and the oscillator is equal to the maximum Rabi frequency for the entire sequence.  The spin's Rabi frequency, phase and detuning are piecewise constant for subpulse durations of either $dt = 0.5T_g$ or $dt=T_g$ and the values during each subpulse are numerically optimized.  In Fig.~(\ref{CINC}), we plot  the fidelity of ${\tt CINC}$, which couples two qudits stored in separate oscillators.  The gate is decomposed into $\cinc'$ and $\bus$ as in Eq.~(\ref{cinc_decomp}).  The spin couples to one oscillator at a time, and is used as a bus between the two oscillators.  We use $\cinc'$ as optimized in Sec.~\ref{fully_numerical_single_mode} and $\bus$ is optimized using the methods of Sec.~\ref{num_opt_sa}.  The fidelities of the two gates are so similar to one another because $\bus$ is synthesized with a very high fidelity. However, a finite time is required to synthesize $\bus$, so $\cinc$ requires more time than $\cinc$'.  
 }
% \label{CINCp_controls}
\end{figure}

We plot the gate fidelity vs. the total time to synthesize the two-qudit $\cinc$ in units of $T_g$ in Fig.~(\ref{CINC}) for $N=2,3$ and for $dt=0.5T_g,T_g$.
Each data point represents a different number of pulses used to synthesize $\cinc'_{1s}$, while for each $N$ and $dt$, $\bus_{s2}$ is uniquely specified as described above.
Because we have chosen to synthesize $\bus_{s2}$ with a relatively high fidelity, the fidelity of $\cinc$ and $\cinc'_{1s}$ are quite close.  On the other hand, $\bus_{s2}$ takes a finite time to synthesize, so the time to synthesize the two-qudit $\cinc$ is somewhat longer than the time to synthesize $\cinc'_{1s}$.  We see that once again for both $N=2,3$ we are able to achieve fidelities on the order of $10^{-4}$.  In conclusion, because the two-qudit $\cinc$ is universal for qudit quantum computation, we have shown that we can synthesize arbitrary multiqudit gates.

\section{Summary}
 We have studied qudit quantum computation in the Jaynes-Cummings model.  Our qudits consist of the first $N$ levels of a harmonic oscillator and our controls consist of the coupling between the oscillator and a spin-1/2 system as well as the drive of the spin.  Beginning with these simple controls, we showed that arbitrary rotations within distinct two level subspaces are possible.  Using these arbitrary rotations, along with the state preparation scheme of Law and Eberly \cite{law_arbitrary_1996}, we designed a protocol to synthesize arbitrary single qudit gates using only resonant interactions between the oscillator and spin.
 While the analytic protocol is sufficient to provide a proof of principle that resonant interactions can be used to synthesize arbitrary transformations,  it was not necessarily optimal in the time it required.  
 However, we were able to use numerical optimization to reduce the time required to synthesize arbitrary transformations.  A comparison between the semi-analytic routine and qubit-based approaches showed comparable times between the two, while our semi-analytic protocol had the advantage of requiring fewer multimode interactions.  Furthermore, because the optimized interactions are relatively simple, the semi-analytic approach can be used with relatively large $N$.
Although the semi-analytic approach provides a substantial speed-up over the analytic protocol, it is still not necessarily optimal.  To speed up the synthesis of qudit transformations even further, we numerically optimized all aspects of the controls, without making any analytic simplifications.  Because of the need to control leakage, this approach is limited to smaller $N$, but can reach high fidelities in a time an order of magnitude shorter than the semi-analytic approach.   Finally, we have shown that extending these protocols to synthesis two-mode gates with high fidelity in a short time is possible.

\begin{acknowledgments}
This work was supported by the EU Integrated project AQUTE, and 
we thank Durga Dasari, Seth Merkel, Paul Pham, and Jake Gulliksen for helpful discussions. 
\end{acknowledgments}

\appendix

\section{Synthesis of $U^{(2)}$ \label{sec:syn_U_2}}

Our goal is to show that transformations of the form Eq.~(\ref{U_2_plain}) are sufficient to synthesize more general transformations of the form Eq.~(\ref{extension_2}).   We cannot synthesize $U^{(2)}$ exactly, so instead we will synthesize an approximation, $U^{(2)}_{a}$ with an error
\begin{equation}
E'^{(2)}= \left| \left| P_C \left( U^{(2)} - U^{(2)}_{a} \right) P_C \right| \right|.
\end{equation}

Where $||M|| = \sqrt{\text{Tr}\{M^\dagger M   \}  }$.  We will also bound the time needed to synthesize $U^{(2)}$ in terms of the qudit dimension, $N+1$, and the desired error, $E'^{(2)}$.  Our method is largely based on \cite{pryor_fourier_2007}, so we present their results for completeness. 

\subsection{Pulse sequence}

We begin by breaking the transformation in Eq.~(\ref{extension_2}) up into it's constituent Euler angles,
\begin{equation}
U^{(2)} = V_{1} V_{2} V_{3}.
\end{equation}
Where
\begin{subequations}
\begin{align}
V_{1} = & \prod_{n=0}^{N+1} e^{i \alpha_{n1} \sigma_{x,n} /2 },\\
V_{2} = &\prod_{n=0}^{N+1} e^{i \alpha_{n2} \sigma_{y,n} /2 },\\
V_{3} = & \prod_{n=0}^{N+1} e^{i \alpha_{n3} \sigma_{x,n} /2 },
\end{align}
\label{V}
\end{subequations}
are the three Euler rotations with $n$ dependent rotation angles.  By construction, $\alpha_{0k}=0$, so we are only dealing with $N+1$ distinct subspaces.  We can perform a discrete cosine transform on the rotation angles \cite{ahmed_discrete_1974},
\begin{equation}
\frac{\alpha_{nk}}{\sqrt{n}} = \sum\limits_{l=0}^{N} a_{kl} \cos\left(  \frac{\pi \left(n+\frac{1}{2} \right) l}{N+1}   \right).
\end{equation}

The Euler rotations may then be written
\begin{equation}
V_{k} = \prod_{l=0}^{N} W_{kl}.
\end{equation}
Where
\begin{align}
W_{1l} = & \prod_{n=0}^{N+1} e^{i  a_{1l} \sqrt{n} \cos\left( \frac{\pi (n+1/2)l}{N+1}    \right) \sigma_{x,n} /2  },\\
W_{2l} = & \prod_{n=0}^{N+1} e^{i  a_{2l} \sqrt{n} \cos\left( \frac{\pi (n+1/2)l}{N+1}  \right) \sigma_{y,n}  /2  },\\
W_{3l} = & \prod_{n=0}^{N+1} e^{i  a_{3l} \sqrt{n} \cos\left(  \frac{\pi (n+1/2)l}{N+1}  \sigma_{x,n} \right)  \sigma_{x,n}   /2  }.
\end{align}
Then $W_{kl}$ corresponds to the $l^{th}$ term in the Fourier series of the $k^{th}$ Euler rotation.  Thus we have reduced the problem of synthesizing the transformations of Eq.~(\ref{extension_2}), whose dependence on the harmonic oscillator level $n$ is arbitrary, to the problem of synthesizing rotations around the $x$ or $y$ axis for which the $n$ dependence is of the form $\sqrt{n}\cos\left(  \frac{\pi (n+1/2)l}{N+1}   \right)$.

To show that the Fourier terms may be synthesized, we begin with a sequence of pulses of the form Eq.~(\ref{U_2_plain})
\begin{align}
t_{a} = & \prod_{k=0}^{N+1}  e^{-i d\phi'  \sqrt{k} \sigma_{x,k} /2  }   \prod_{l=0}^{N+1} e^{-i d\phi'  \sqrt{l} \sigma_{y,l}/2   }
 \prod_{m=0}^{N+1} e^{i d\phi'  \sqrt{m} \sigma_{x,m} /2  }   \prod_{n=0}^{N+1}  e^{i d\phi'  \sqrt{n} \sigma_{y,n_p}  /2 } \nonumber\\
  & \approx  \prod_{n=0}^{N+1} e^{-i\, n\,d\phi\,\sigma_{z,n}/2}.
\label{t_a}
\end{align}
Where $d\phi = d\phi'^2$ is small enough the above approximation holds.  Repeating this procedure $Q$ times allows us to generate larger rotations around the $z$-axis,
\begin{equation}
T_{a} = t^Q \approx \prod_{n=0}^{N+1} e^{-i\, n \,\phi\,\sigma_{z,n}/2}.
\label{T_a}
\end{equation}
Where $\phi/Q = d\phi$  and $\phi$ is not necessarily small.  For Eq.~(\ref{T_a}) to hold, $Q$ must be sufficiently large and we will determine how large below.   If we choose $\phi = 2\pi l/N$, then we can synthesize.
\begin{subequations}
\begin{align}
w_{a,kl1} = & T_{a}^\dagger \tilde{U}_2 \left( g_{max},0,\frac{l\pi}{N+1},\frac{d\theta_{kl}  }{g_{max} }\right) T_{a}   \nonumber \\
\approx & \prod_{n=0}^{N+1}  e^{-i d\theta_{kl}  \sqrt{n} \left( \cos(   \frac{\pi (n+1/2)l}{N+1}    ) \sigma_{x,n}  - \sin(  \frac{\pi (n+1/2)l}{N+1}  ) \sigma_{y,n} \right)/2  }, \\
w_{a,kl2} = & T_{a} \tilde{U}_2 \left( g_{max},0,-\frac{l\pi}{N+1},\frac{d\theta_{kl}  }{g_{max} }\right)  T_{a}^\dagger  \nonumber \\
\approx &\prod_{n=0}^{N+1}  e^{-i d\theta_{kl}  \sqrt{n} \left( \cos(  \frac{\pi (n+1/2)l}{N+1}  ) \sigma_{x,n}  + \sin( \frac{\pi (n+1/2)l}{N+1}  ) \sigma_{y,n} \right)/2  }.
\end{align}
\label{w_a}
\end{subequations}
For $d\theta_{kl}$ sufficiently small, we then have
\begin{equation}
w_{a,kl1}w_{a,kl2}\approx \prod_{n=0}^{N+1}  e^{-i 2 d\theta_{kl}  \sqrt{n} \cos(  \frac{\pi (n+1/2)l}{N+1}  ) \sigma_{x,n}  /2  }.
\label{prod_w_a_kl1_w_a_kl2}
\end{equation}
Repeating the above sequence $P$ times allows us to synthesize the net transformation
\begin{equation}
W_{a, kl} = (w_{a,kl1} w_{a,kl2})^P \approx \prod_{n=0}^{N+1}  e^{-i \alpha_{kl}  \sqrt{n}  \cos(  \frac{\pi (n+1/2)l}{N+1}  ) \sigma_{x,n}  /2  }.
\label{W_a_kl}
\end{equation}
Where $d\theta_{kl} = \alpha_{kl}/2P$.  For $P$ sufficiently large, we have $W_{a,kl} \approx W_{kl}$.  Since this shows we can generate arbitrary terms in the Fourier expansion of the Euler angles which define the rotation in Eq.~(\ref{extension_2}),  we have proven that we can synthesize arbitrary transformations of the form, Eq.~(\ref{extension_2}).

\subsection{Relating $P$ and $Q$ to errors synthesizing $U^{(2)}$}
In this section, we relate the error in synthesizing $U^{(2)}$ to $P$ and $Q$.  This relation will allow us to discuss the trade off between the accuracy and time with which we synthesize $U^{(2)}$ in the next section.
Direct calculations show us that if
\begin{equation}
T = \prod_{n=0}^{N+1} e^{-i\, n \,\phi\,\sigma_{z,n}/2}.
\end{equation}
then the error in synthesizing $T_{a}$ is
\begin{subequations}
\begin{align}
\epsilon'_T = & \left| \left| P_C \left( T - T_{a} \right) P_C \right| \right| \\
\leq  & \frac{   4 \left(  2\pi  \right)^{3/2}     \left(N+1 \right) ^{5/2}   }{ \sqrt{Q}}.
\end{align}
\end{subequations}
Where we have neglected terms of order $\frac{(N+1)^3}{Q}$ and smaller.

Furthermore, the error in synthesizing any particular $W_{kl}$ is
\begin{equation}
\epsilon'_{W_{kl}} = \left| \left|P_C \left( W_{kl} - W_{a,kl} \right) P_C \right| \right|
		\leq  P \left| \left| P_C \left( W_{kl}^{1/p} - w_{a,kl1} w_{a,kl2}  \right) P_C \right|\right| .\\
\end{equation}
If we define
\begin{subequations}
\begin{align}
w_{kl1} = & T^\dagger \tilde{U}_2 \left( g_{max},0,\frac{l\pi}{N+1},\frac{d\theta_{kl}  }{g_{max} }\right) T   \nonumber \\
= & \prod_{n=0}^{N+1}  e^{-i d\theta_{kl}  \sqrt{n} \left( \cos(   \frac{\pi (n+1/2)l}{N+1}    ) \sigma_{x,n}  - \sin(  \frac{\pi (n+1/2)l}{N+1}  ) \sigma_{y,n} \right)/2  }, \\
w_{kl2} = & T \tilde{U}_2 \left( g_{max},0,-\frac{l\pi}{N+1},\frac{d\theta_{kl}  }{g_{max} }\right)  T^\dagger  \nonumber \\
= &\prod_{n=0}^{N+1}  e^{-i d\theta_{kl}  \sqrt{n} \left( \cos(  \frac{\pi (n+1/2)l}{N+1}  ) \sigma_{x,n}  + \sin( \frac{\pi (n+1/2)l}{N+1}  ) \sigma_{y,n} \right)/2  }.
\end{align}
\end{subequations}
then the error in $w_{a,klj} $ is
\begin{equation}
\epsilon'_{w_{a,klj} } = \left| \left| P_C \left(   w_{klj}   - w_{a,klj}  \right)P_C \right| \right|.
\end{equation}
Since  $ \epsilon'_{w_{a,kl1} } \approx  \epsilon'_{w_{a,kl2} } $ we can bound $\epsilon'_{W_{kl}}$
\begin{equation}
\epsilon'_{W_{kl}} \leq P\left( \left| \right|P_C\left( W^{1/p}_{kl} - w_{kl1} w_{kl2}  \right) P_C \left| \right| + 2 \epsilon'_{w_{a,kl1} }  \right).
\end{equation}
Furthermore,
\begin{equation}
\epsilon'_{w_{a,klj} }  \leq 2 \epsilon'_T.
\end{equation}
and
\begin{equation}
\left| \left| P_C \left( W_{kl}^{1/p} - w_{kl1} w_{kl2}  \right) P_C \right| \right| \leq \sqrt{2} \left( \frac{2\pi}{P}   (N+1)  \right)^2.
\end{equation}
Where we have neglected terms of order $\frac{(N+1)^{5/2}}{P^3}$ and smaller. Then
\begin{equation}
\epsilon'_{W_{kl}} \leq P \left(\sqrt{2} \left( \frac{2\pi}{P} (N+1) \right)^2   + 16\frac{( 2\pi)^{3/2} (N+1)^{5/2} }{\sqrt{Q}} \right).
\label{epsilon_W_kl}
\end{equation}

Now we relate the errors  in the individual Fourier terms of the Euler angles, $W_{a,kl}$, to the errors in $U^{(2)}_{a}$,
\begin{equation}
E'^{(2)} \leq   3(N+1) \max_{k,l} \left| \right| P_C \left( W_{kl} - W_{kl,a} \right) P_C \left| \right|.
 \end{equation}
Since the bound on $\left| \right| P_C \left( W_{kl} - W_{kl,a} \right) P_C\left|\right|$, Eq.~(\ref{epsilon_W_kl}), is independent of $k$ and $l$,  we have
\begin{equation}
E'^{(2)}  \leq  3P \left(\sqrt{2}(N+1)^3 \left( \frac{2\pi}{P}  \right)^2   + 16\frac{( 2\pi)^{3/2} (N+1)^{7/2} }{\sqrt{Q}} \right).
\label{epsilon_2}
\end{equation}

\subsection{Time to synthesize $U^{(2)}$}
Let $T(M)$ be the time to synthesize a transformation $M$, then
\begin{equation}
T(U^{(2)}_{a}) \leq 3(N+1) \max_{kl}\, T(W_{a,kl}).
\end{equation}
From Eq.~(\ref{W_a_kl}) and Eq.~(\ref{w_a}) we see that
\begin{equation}
T(W_{a,kl}) = 4P T(T_{a}).
\end{equation}
The time to synthesize $T_a$ is bounded by $T(T_a) \leq 4 T_g \sqrt{\frac{Q}{2\pi}}$
so that $T(W_{a,kl}) =\frac{16}{\sqrt{2\pi}}P \sqrt{Q} T_g$, where $T_g = \frac{2\pi}{g}$.
Then
\begin{equation}
T(U^{(2)}_{a}) \leq \frac{48}{\sqrt{2\pi}} P \sqrt{Q} T_g  (N+1).
\label{time_P_Q}
\end{equation}

We are free to choose $P$ and $Q$, but comparing Eq.~(\ref{epsilon_2}) and Eq.~(\ref{time_P_Q}), we see there is a trade off between minimizing $E'^{(2)}$ and $T(U^{(2)})$.   For most cases, we have a target accuracy, $E'^{(2)}$, and we would like to minimize the time required to reach that accuracy.  We do so by first  choosing $P$ and $Q$ so that equality holds in Eq.~(\ref{epsilon_2}), which gives
\begin{equation}
Q = \frac{18432 \pi^3 (N+1)^7 P^4}{\left(  12\sqrt{2} \pi^2 (N+1)^3 - E'^{(2)}P  \right)^2}.
\end{equation}
Plugging this value for $Q$ into Eq.~(\ref{time_P_Q}) and finding the value of $P$ which minimizes $T(U^{(2)}_{a})$, gives
\begin{equation}
T(U^{(2)}_{a}) \leq k^{(2)} T_g \frac{(N+1)^{10.5}}{E'^{(2)3}} .
\label{time_U_2_a}
\end{equation}
Where $k^{(2)} \approx 2.7 \times 10^{9}$.

There are several reason why the scaling with $N$ and $E^{(a)}$ is so bad.  The first problems arise from the approximation used to synthesize $t_a$.  Even though $t_a$ is effectively a very small rotation, the time to synthesize it is still proportional to $1/\sqrt{Q}$.
Then $t_a$ is repeated $Q$ times to synthesize $T_a$ and in order for the approximation Eq.~(\ref{t_a}) to hold, $Q$ must be  large, and so synthesizing $T_a$ requires a time much larger than $T_g$.  To make matters worse, $T_a$ is used to synthesize the effectively infinitesimal pulses, $w_{klj,a}$.  Because $w_{klj,a}$ is repeated  $P$ times to synthesize $W_{kl,a}$, we must repeat $T_a$.  In order for the approximation in Eq.~(\ref{prod_w_a_kl1_w_a_kl2}) to hold, $P$ must be  large and because we use $T_a$  so many times, it must be  synthesized with extremely high precision. In addition, because $T(w_{klj,a})$ is finite and $w_{klj,a}$ is repeated $P$ times,  $T(W_{kl,a})$ is extremely large, which in turn forces our estimate for $T(U^{(2)})$ to be extremely large.  Thus, the combination of the approximations in Eq.~(\ref{t_a}) and Eq.~(\ref{prod_w_a_kl1_w_a_kl2}), and the finite time required to synthesize $t_a$ conspire to generate terrible scaling with $N$ and $E^{(a)}$.  We conclude by noting that $u^{(2)}_{j\N}$ is a special case of $U^{(2)}$, so we have also shown in this section that we can synthesize $u^{(2)}_{j\N}$.

\section{Synthesis of $u^{(1)}_{j\N}$  \label{syn_u_j_jN} }
Next, we construct transformations of the form Eq.~(\ref{first_u_1_j}). We begin by noting that we have already shown we can synthesize transformations of the form Eq.~(\ref{V_1_n}), because they have the same form as Eq.~(\ref{extension_2}).  Thus to construct $u^{(1)}_{j\N}$ we can use either Eq.~(\ref{u_1_n_syn}) or Eq.~(\ref{u_1_n_syn_z}), depending on whether the torque vector of $M^{(1)}_j$ has a $z$-component.

The time required to synthesize $u^{(1)}_{j\N,a}$
is dominated by the time to synthesize $V^{(1)}_{n\N,a}$.  Since the bound on $T\left(V^{(1)}_{n\N,a}\right )$ is given by  Eq.~(\ref{time_U_2_a}) we have
\begin{equation}
T(u^{(1)}_{j\N,a}) \leq 2 k^{(2)} T_g \frac{(N+1)^{10.5}}{E'^{(2)3}}.
\end{equation}
However, we would prefer to relate this to the accuracy with which we synthesize $u^{(1)}_{j\N,a}$, $\epsilon'^{(1)}$. 
Since the errors in synthesizing $u^{(1)}_{j\N,a}$ are due entirely to the errors in synthesizing $V^{(1)}_{n\N,a}$,  we have $\epsilon'^{(1)} =2 E'^{(2)}$, and so
\begin{equation}
T(u^{(1)}_{j\N,a}) \leq k^{(1)} T_g \frac{(N+1)^{10.5}}{\epsilon'^{(1)3}}.
\end{equation}
Where $k^{(1)} \approx 4.4 \times 10^{10}$.

\section{Treating the overall phase}

In many cases, we are not concerned about the overall phase of the transformation we seek to synthesize and so we are usually interested in minimizing
\begin{equation}
\epsilon^{(a)}= \min_{\phi} \left| \left|P_C \left( U^{(a)}- e^{i\phi}U^{(a)}_{a}  \right) P_C\right|\right|.
\end{equation}
With $a= 1,2$.
However neither the error in $T_{a}$ or $W_{kl}$ is improved by ignoring the overall phase,
\begin{subequations}
\begin{align}
\min_{\phi} \left| \left|P_C  \left(T- e^{i\phi}T_{a} \right) P_C \right| \right| = & \left| \left|P_C \left(T- T_{a} \right) P_C  \right|\right|, \\
\min_{\phi} \left| \left| P_C \left( W_{kl}^{1/p} - e^{i\phi}w_{kl1} w_{kl2}  \right) P_C \right| \right| = & \left| \left| P_C \left( W_{kl}^{1/p} - w_{kl1} w_{kl2}\right) P_C\right| \right|.
\end{align}
\end{subequations}
As a result, ignoring the overall phase does not improve the error or shorten the time to synthesize a pulse when using the analytic construction of the appendix.  Thus, we can replace $\epsilon'^{(a)}$ with $\epsilon^{(a)}$ and the calculated times remain unchanged.
For comparison with the numerical optimization presented later we give the time to synthesize a pulse in terms of  $\epsilon^{(a)}$,
\begin{subequations}
\begin{align}
T(U^{(2)}_{a}) \leq &k^{(2)} T_g \frac{(N+1)^{10.5}}{\epsilon^{(2)3}},  \\
T(u^{(1)}_{j\N,a}) \leq & k^{(1)} T_g \frac{(N+1)^{10.5}}{\epsilon^{(1)3}}.
\end{align}
\end{subequations}

\section{Time to synthesize arbitrary transformations}

Now we are ready to give the time required to synthesize an arbitrary unitary transformation on the computational space.  In Sec.~\ref{analytic} we described how to synthesize an arbitrary transformation $U$ using $u^{(a)}_{j\N}$.  Since we now have a method to synthesize $u^{(a)}_{j\N,a}\approx u^{(a)}_{j\N}$  we can use  $u^{(a)}_{j\N,a}$ to synthesize a transformation $U_a$ that approximates our target transformation, $U$, with an error bounded by $\delta$,
\begin{equation}
\min_{\phi} \left| \left| P_C \left( U - e^{i\phi}U_{a}  \right) P_C \right| \right| \leq \delta.
\end{equation}
As described in App.~\ref{syn_u_j_jN}, $u^{(1)}_{j\N,a}$ is synthesized using several $u^{(2)}_{j\N,a}$'s. As a result, all the errors which result in synthesizing $U_a$ are due to the errors in synthesizing $u^{(2)}_{j\N,a}$.
If $g_{a}(N) = 3N^2+5N+2$ is the number of $u^{(2)}_{j\N,a}$ used to construct $U_{a}$, then the error with which we synthesize any given $u^{(2)}_{j\N,a}$ must be bounded by
\begin{equation}
\epsilon(u^{(2)}_{j\N,a}) \leq \frac{\delta}{g_{a}(N)}.
\end{equation}
Then the total time for one of the $u^{(2)}_{j\N,a}$'s used to synthesize $U_a$ is
\begin{equation}
T(u^{(2)}_{j\N,a}) \leq k^{(2)} T_g \frac{(3N^2+5N+2)^3(N+1)^{10.5}}{\delta^3}.
\end{equation}
Finally, because the time to synthesize $U_a$ is dominated by the time to synthesize $u^{(2)}_{j\N,a}$ and there are $g_a(N)$ such transformations needed to synthesize $U_a$, the total time is
\begin{equation}
T(U_{a}) \leq k^{(2)} T_g \frac{(3N^2+5N+2)^4(N+1)^{10.5}}{\delta^3}.
\end{equation}
It's also common to deal with a quantity which can be related to the gate fidelity, \cite{motzoi_optimal_2011}, 
\begin{equation}
\eta = \min_{\phi} \frac{1}{4(N+1)}\left| \left| P_C \left( U - e^{i\phi}U_{a} \right) P_C \right| \right|^2 .
\end{equation}
In terms of $\eta$ the total time is
\begin{equation}
T(U_{a})  \leq k T_g \frac{(3N^2+5N+2)^4(N+1)^{9}}{\eta^{3/2}} .
\label{time_analytic}
\end{equation}
Where $k \approx 3.4\times10^{8}$.

\bibliographystyle{apsrev4-1}

\bibliography{jcm_qudit_v19_2}

\end{document}